\documentclass[authorversion,nonacm,sigconf,natbib]{acmart}

\usepackage{adjustbox}
\usepackage{csquotes}
\usepackage{multirow}
\usepackage[caption=false,font=footnotesize]{subfig}

\usepackage[capitalise,nameinlink]{cleveref}
\crefname{lstlisting}{\lstlistingname}{\lstlistingname}
\Crefname{lstlisting}{Listing}{Listings}

\usepackage{hyperref}

\newcommand{\eg}{ e.\,g.,\ }
\newcommand{\ie}{ i.\,e.,\ }

\newcommand{\mrotrow}[2]{\parbox[t]{1mm}{\multirow{#1}{*}{\rotatebox[origin=c]{90}{#2}}}}

\usepackage{todonotes}

\newcommand{\IMsize}{559}
\newcommand{\IMconn}{63\,973\,325}
\newcommand{\IMalerts}{12\,735}
\newcommand{\IMweakalerts}{4\,385}

\newcommand{\IMgacalerts}{125}

\newcommand{\IMhighlevelalerts}{4\,510}
\newcommand{\IMinfectiongraphs}{511}
\newcommand{\IMtotalscenarios}{4\,418}
\newcommand{\IMdistinctscenarios}{642}

\newcommand{\IMreductionpc}{5.04}
\newcommand{\IMscenariosperday}{34.2}

\newcommand{\IFalerts}{446\,458}
\newcommand{\IFweakalerts}{39\,754}

\newcommand{\IFgacalerts}{10\,724}

\newcommand{\IFhighlevelalerts}{50\,478}
\newcommand{\IFinfectiongraphs}{491}
\newcommand{\IFtotalscenarios}{4\,253}
\newcommand{\IFdistinctscenarios}{700}

\newcommand{\IFreductionpc}{0.16}
\newcommand{\IFscenariosperday}{70}

\newcommand{\AMalerts}{12\,675}
\newcommand{\AMweakalerts}{4\,432}

\newcommand{\AMgacalerts}{119}

\newcommand{\AMhighlevelalerts}{4\,551}
\newcommand{\AMinfectiongraphs}{442}
\newcommand{\AMtotalscenarios}{3\,452}
\newcommand{\AMdistinctscenarios}{611}

\newcommand{\AMreductionpc}{4.82}

\newcommand{\AFalerts}{446\,407}
\newcommand{\AFweakalerts}{39\,649}

\newcommand{\AFgacalerts}{10\,713}

\newcommand{\AFhighlevelalerts}{50\,362}
\newcommand{\AFinfectiongraphs}{456}
\newcommand{\AFtotalscenarios}{4\,305}
\newcommand{\AFdistinctscenarios}{686}

\newcommand{\AFreductionpc}{0.15}
\newcommand{\AFscenariosperday}{68.6}

\newcommand{\AMatkexec}{---}
\newcommand{\AMatklm}{---}

\newcommand{\AMatklmextract}{---}

\newcommand{\AMconnre}{1\,171}
\newcommand{\AMjswebinj}{336}
\newcommand{\AMlargetx}{5\,772}
\newcommand{\AMmultilargetx}{187}
\newcommand{\AMsensitivesig}{---}
\newcommand{\AMsighttp}{6}
\newcommand{\AMsignonhttp}{4}
\newcommand{\AMsmbtransfer}{1}
\newcommand{\AMsqlinj}{78}
\newcommand{\AMstalledhttp}{4\,976}
\newcommand{\AMverylargetx}{10}
\newcommand{\AMweblogin}{14}
\newcommand{\AMdouplepulsar}{---}
\newcommand{\AMetblue}{---}
\newcommand{\AMetsyn}{---}
\newcommand{\AMviocmd}{---}
\newcommand{\AMviontrename}{---}
\newcommand{\AMviopidmid}{---}
\newcommand{\AMviotxcmd}{---}

\newcommand{\AMweakkey}{120}

\newcommand{\AMatkexectp}{---}
\newcommand{\AMatklmtp}{---}

\newcommand{\AMatklmextracttp}{---}

\newcommand{\AMconnretp}{0}
\newcommand{\AMjswebinjtp}{0}
\newcommand{\AMlargetxtp}{0}
\newcommand{\AMmultilargetxtp}{0}
\newcommand{\AMsensitivesigtp}{---}
\newcommand{\AMsighttptp}{0}
\newcommand{\AMsignonhttptp}{1}
\newcommand{\AMsmbtransfertp}{1}
\newcommand{\AMsqlinjtp}{0}
\newcommand{\AMstalledhttptp}{0}
\newcommand{\AMverylargetxtp}{1}
\newcommand{\AMweblogintp}{0}
\newcommand{\AMdouplepulsartp}{---}
\newcommand{\AMetbluetp}{---}
\newcommand{\AMetsyntp}{---}
\newcommand{\AMviocmdtp}{---}
\newcommand{\AMviontrenametp}{---}
\newcommand{\AMviopidmidtp}{---}
\newcommand{\AMviotxcmdtp}{---}

\newcommand{\AMweakkeytp}{0}
\newcommand{\AMtotaltp}{3}

\newcommand{\AMatkexectpr}{---}
\newcommand{\AMatklmtpr}{---}

\newcommand{\AMatklmextracttpr}{---}

\newcommand{\AMconnretpr}{0.00}
\newcommand{\AMjswebinjtpr}{0.00}
\newcommand{\AMlargetxtpr}{0.00}
\newcommand{\AMmultilargetxtpr}{0.00}
\newcommand{\AMsensitivesigtpr}{---}
\newcommand{\AMsighttptpr}{0.00}
\newcommand{\AMsignonhttptpr}{0.25}
\newcommand{\AMsmbtransfertpr}{1.00}
\newcommand{\AMsqlinjtpr}{0.00}
\newcommand{\AMstalledhttptpr}{0.00}
\newcommand{\AMverylargetxtpr}{0.10}
\newcommand{\AMweblogintpr}{0.00}
\newcommand{\AMdouplepulsartpr}{---}
\newcommand{\AMetbluetpr}{---}
\newcommand{\AMetsyntpr}{---}
\newcommand{\AMviocmdtpr}{---}
\newcommand{\AMviontrenametpr}{---}
\newcommand{\AMviopidmidtpr}{---}
\newcommand{\AMviotxcmdtpr}{---}

\newcommand{\AMweakkeytpr}{0.00}
\newcommand{\AMtotaltpr}{0.000237}
\newcommand{\AMtotaltprpc}{0.0237}

\newcommand{\AFatkexec}{2}
\newcommand{\AFatklm}{4}

\newcommand{\AFatklmextract}{1}

\newcommand{\AFconnre}{\AMconnre}
\newcommand{\AFjswebinj}{\AMjswebinj}
\newcommand{\AFlargetx}{\AMlargetx}
\newcommand{\AFmultilargetx}{\AMmultilargetx}
\newcommand{\AFsensitivesig}{8\,731}
\newcommand{\AFsighttp}{\AMsighttp}
\newcommand{\AFsignonhttp}{\AMsignonhttp}
\newcommand{\AFsmbtransfer}{\AMsmbtransfer}
\newcommand{\AFsqlinj}{\AMsqlinj}
\newcommand{\AFstalledhttp}{\AMstalledhttp}
\newcommand{\AFverylargetx}{\AMverylargetx}
\newcommand{\AFweblogin}{\AMweblogin}
\newcommand{\AFdouplepulsar}{1}
\newcommand{\AFetblue}{53}
\newcommand{\AFetsyn}{1}
\newcommand{\AFviocmd}{1\,389}
\newcommand{\AFviontrename}{8\,731}
\newcommand{\AFviopidmid}{6\,133}
\newcommand{\AFviotxcmd}{408\,686}

\newcommand{\AFweakkey}{\AMweakkey}

\newcommand{\AFatkexectp}{2}
\newcommand{\AFatklmtp}{4}

\newcommand{\AFatklmextracttp}{1}

\newcommand{\AFconnretp}{\AMconnretp}
\newcommand{\AFjswebinjtp}{\AMjswebinjtp}
\newcommand{\AFlargetxtp}{\AMlargetxtp}
\newcommand{\AFmultilargetxtp}{\AMmultilargetxtp}
\newcommand{\AFsensitivesigtp}{0}
\newcommand{\AFsighttptp}{\AMsighttptp}
\newcommand{\AFsignonhttptp}{\AMsignonhttptp}
\newcommand{\AFsmbtransfertp}{\AMsmbtransfertp}
\newcommand{\AFsqlinjtp}{\AMsqlinjtp}
\newcommand{\AFstalledhttptp}{\AMstalledhttptp}
\newcommand{\AFverylargetxtp}{\AMverylargetxtp}
\newcommand{\AFweblogintp}{\AMweblogintp}
\newcommand{\AFdouplepulsartp}{1}
\newcommand{\AFetbluetp}{0}
\newcommand{\AFetsyntp}{1}
\newcommand{\AFviocmdtp}{0}
\newcommand{\AFviontrenametp}{0}
\newcommand{\AFviopidmidtp}{0}
\newcommand{\AFviotxcmdtp}{1}

\newcommand{\AFweakkeytp}{\AMweakkeytp}
\newcommand{\AFtotaltp}{13}

\newcommand{\AFatkexectpr}{1.00}
\newcommand{\AFatklmtpr}{1.00}

\newcommand{\AFatklmextracttpr}{1.00}

\newcommand{\AFconnretpr}{\AMconnretpr}
\newcommand{\AFjswebinjtpr}{\AMjswebinjtpr}
\newcommand{\AFlargetxtpr}{\AMlargetxtpr}
\newcommand{\AFmultilargetxtpr}{\AMmultilargetxtpr}
\newcommand{\AFsensitivesigtpr}{0.00}
\newcommand{\AFsighttptpr}{\AMsighttptpr}
\newcommand{\AFsignonhttptpr}{\AMsignonhttptpr}
\newcommand{\AFsmbtransfertpr}{\AMsmbtransfertpr}
\newcommand{\AFsqlinjtpr}{\AMsqlinjtpr}
\newcommand{\AFstalledhttptpr}{\AMstalledhttptpr}
\newcommand{\AFverylargetxtpr}{\AMverylargetxtpr}
\newcommand{\AFweblogintpr}{\AMweblogintpr}
\newcommand{\AFdouplepulsartpr}{1.00}
\newcommand{\AFetbluetpr}{0.00}
\newcommand{\AFetsyntpr}{1.00}
\newcommand{\AFviocmdtpr}{0.00}
\newcommand{\AFviontrenametpr}{0.00}
\newcommand{\AFviopidmidtpr}{0.00}
\newcommand{\AFviotxcmdtpr}{0.000002}

\newcommand{\AFweakkeytpr}{\AMweakkeytpr}
\newcommand{\AFtotaltpr}{0.00000245}
\newcommand{\AFtotaltprpc}{0.000245}

\begin{document}

\title[Multi-Stage Attack Detection via Kill Chain State Machines]{Multi-Stage Attack Detection via Kill Chain State Machines}

%
\author{Florian Wilkens}
\affiliation{%
  \institution{Universität Hamburg}
  \state{Hamburg}
  \country{Germany}
}
\email{wilkens@informatik.uni-hamburg.de}

\author{Felix Ortmann}
\affiliation{%
  \institution{Tenzir GmbH}
  \state{Hamburg}
  \country{Germany}
}
\email{felix.ortmann@tenzir.com}

\author{Steffen Haas}
\affiliation{%
  \institution{Universität Hamburg}
  \state{Hamburg}
  \country{Germany}
}
\email{haas@informatik.uni-hamburg.de}

\author{Matthias Vallentin}
\affiliation{%
  \institution{Tenzir GmbH}
  \state{Hamburg}
  \country{Germany}
}
\email{matthias.vallentin@tenzir.com}

\author{Mathias Fischer}
\affiliation{%
  \institution{Universität Hamburg}
  \state{Hamburg}
  \country{Germany}
}
\email{mfischer@informatik.uni-hamburg.de}

\renewcommand{\shortauthors}{Wilkens and Ortmann, et al.}

\begin{abstract}
Today, human security analysts collapse under the sheer volume of alerts they have to triage during
investigations. The inability to cope with this load, coupled with a high false positive rate of
alerts, creates \emph{alert fatigue}. This results in failure to detect complex attacks, such as
advanced persistent threats~(APTs), because they manifest over long time frames and attackers tread
carefully to evade detection mechanisms.
In this paper, we contribute a new method to synthesize attack graphs from state machines. We use
the network direction to derive potential attack stages from single and meta-alerts and model
resulting attack scenarios in a \emph{kill chain state machine~(KCSM)}. Our algorithm yields a
graphical summary of the attack, \emph{APT scenario graphs}, where nodes represent involved hosts
and edges infection activity.
We evaluate the feasibility of our approach in multiple experiments based on the CSE-CIC-IDS2018
data set~\cite{ids-sharafaldin-2018}. We obtain up to \IFalerts\ singleton alerts that our
algorithm condenses into \IFdistinctscenarios~APT scenario graphs resulting in a reduction of up to
three orders of magnitude. This reduction makes it feasible for human analysts to effectively
triage potential incidents. An evaluation on the same data set, in which we embedded a synthetic
yet realistic APT campaign, supports the applicability of our approach of detecting and
contextualizing complex attacks. The APT scenario graphs constructed by our algorithm correctly
link large parts of the APT campaign and present a coherent view to support the human analyst in
further analyses.
\end{abstract}



\maketitle

\section{Introduction}\label{sec:intro}
The proliferation of security monitoring and mainstream adoption of intrusion detection
systems~(IDS) has shifted the bottleneck during incident response from a lack of visibility to a
lack of human analysis capacity. The massive volume of incoming alerts overwhelms security teams,
who already suffer from chronic under-staffing. Coupled with a high false positive rate,
\emph{alert fatigue} creeps into daily operations and operators can no longer effectively separate
the signal from the noise.
Especially complex long-running attacks conducted by skilled attackers, also referred to as
advanced persistent threats~(APT) might remain undetected. This is a major problem because these
attacks typically inflict financial damage in the double-digit million
dollars~\cite{accenture-2019}. A central research question must therefore lie in developing
understanding and mechanism to detect security incidents \emph{before} they turn into a major
breach.

Numerous approaches to model attack behavior of multi-stage attacks exist in the
literature~\cite{model-siddiqi-2016,model-hutchins-2011,model-hahn-2015}. The unified kill chain
(UKC)~\cite{model-pols-2017} evolved as the most comprehensive and prominent model to differentiate
individual APT stages. However, to the best of our knowledge, there is no approach that fully
implements this model nor any of its predecessors like the intrusion kill chain
(IKC)~\cite{model-hutchins-2011}. Existing approaches either just detect a single stage or only
some connected stages, so that complex attacks cannot be fully detected. Furthermore, the spatial
and temporal distribution of alerts related to complex attacks impedes their detection via
conventional intrusion detection methods that usually operate in a batch-based fashion.

In this paper, we propose a new approach to make alerts more actionable, so that analysts spend
less time triaging. We do so by first using an established alert correlation approach to preprocess
the alert corpus into clustered meta-alerts and unclustered alerts. Next, we leverage the network
direction to assign potential attack stages to all alerts and meta-alerts. Using further
correlation and contextualization based on a \emph{kill chain state machine~(KCSM)} we synthesize
\emph{APT scenario graphs} that describe the individual steps of the multi-stage attack at the
machine boundary.
A unique feature of our approach is the use of topology information to contextualize the analysis.
The typical segmentation of a network into zones makes it possible to infer directionality in the
communication and enrich alerts with this information to infer potential attack stages according to
the KCSM. In the next step we connect the alerts based on these potential stages in a graph and
detect paths that resemble valid transitions according to the ordering of kill chain states in the
KCSM. Ultimately, we obtain scenario graphs that provide the much-needed reduction of the original
alert set and offer additional context about the potential APT campaign to the human analyst.

Our end-to-end prototype first distills raw network traffic into detailed protocol summaries with
the Zeek IDS~\cite{paxson1999bro} and then uses graph-based alert
correlation~(GAC)~\cite{ac-haas-2018} to obtain higher-level alerts. These alerts are processed in
batches by our Killchain-based APT contextualization to construct attack scenario graphs by
maintaining a resource-efficient and compressed representation of all prior batches and their
alerts. This allows also to detect long-running APT attacks without the need to process all alerts
at once together.
We evaluate our prototype on realistic data in two different ways. First, we apply it to the raw
CSE-CIC-IDS2018 data set~\cite{ids-sharafaldin-2018}, where our approach yields a two orders of
magnitude reduction in alert volume from \IFalerts\ alerts to \IFdistinctscenarios\ scenario
graphs. In the absence of ground truth, we have to classify them as false positives.
We then inject a real-world multi-stage attack spanning ten days into the data set and demonstrate
that our algorithm identifies the attack among \AFdistinctscenarios\ generated scenario graphs or
\AFscenariosperday\ scenarios per day. For high-security environments like a government or critical
infrastructure this number should already be manageable by human analysts. Furthermore, the set of
resulting scenarios can be prioritized or filtered\eg{}if critical hosts are involved or additional
unrelated indicators of compromise are available.

The remainder of this paper is structured as follows. We review in~\cref{sec:related_work} related
work on alert correlation, APT models, and model-based APT detection mechanisms.
In~\cref{sec:system} we describe two variations of the \emph{kill chain state machine~(KCSM)} and
our approach to contextualize multi-stage attacks based on it. We evaluate our approach
in~\cref{sec:eval} with respect to accuracy and performance, and outline how KCSM supports
real-world incident response scenarios. In \cref{sec:conclusion} we conclude with a summary and
an outlook.

\section{Related Work}\label{sec:related_work}
\subsection{Alert Correlation and Volume Reduction}\label{subsec:related_work-alert_correlation}
Julisch proposes to cluster alerts according to their root cause~\cite{ac-julisch-2003}. However,
root causes cannot be exactly determined and must be approximated. Therefore, alerts are grouped
based on attribute similarity, which is also known as attribute oriented induction~(AoI).
Zhou et al.\ propose a different similarity-based clustering approach~\cite{ac-zhou-2009}. They
leverage alert attributes as well, but matching attribute sets are predefined in lattice
structures. Though highly efficient, this approach is limited to a priori known patterns.
Zhang et al.\  propose a clustering strategy based on rough sets~\cite{ac-zhang-2018}. A projection
function determines alert equivalence, instead of comparing alert attributes directly. Alert
projections are correlated when they exhibit a certain timely proximity.
Another noteworthy clustering strategy is based on alert entropy~\cite{ac-ghasemigol-2015}. The
authors observe that more frequent events carry lesser information than infrequent events. Partial
entropy is used to identify alert groups with similar information content.

Each of these clustering approaches identifies attacks as a group of alerts. However, long running
and stealthy attacks may not be represented as a \emph{group} of alerts. Existing approaches
operate on batches or temporal proximity of events. Depending on batch size, temporal dispersion of
related alerts, and other factors, these approaches may struggle to correctly correlate alerts that
belong to the extensive attacks often present in APT scenarios.

\subsection{Attack Story Reconstruction}\label{subsec:related_work-attack_reconstruction}
Most alert clustering approaches for reconstruction of attack stories operate in at least two
steps: identify individual attack steps and then connect the steps for a holistic view on an attack
scenario. Strategies for the second step differ with each approach, ranging from causal
connections~\cite{ac-ning-2002,ac-ning-2010}, to mathematical, or graph-based
strategies~\cite{ac-cuppens-2002,ac-fredj-2015}.

Daneshgar and Abbaspour present such a two-fold scenario
reconstruction~\cite{ac-faraji-daneshgar-2016}. IDS alerts are clustered to identify attacks. The
clustering step considers conventional attribute similarity together with a fuzzy, causal
relationship indicator called correlation strength. Lastly, the system incorporates timely
proximity of alerts before clustering. In a second step, the authors use a fuzzy variation of
frequent pattern mining to interconnect these extracted attack steps.
Similarly, Haas and Fischer detect multi-step attacks by first clustering alerts and then
interconnecting those clusters~\cite{ac-haas-2018}. The initial clustering is based on partial
alert attribute similarity. As with many clustering approaches, the authors use a configurable
minimum cluster size to discard non-relevant alert groups.

Many multi-stage attack correlation approaches exhibit the same general problem as pure alert
clustering approaches. Most research focuses on the second step for attack interconnection.
However, the first step for attack identification often uses some sort of alert clustering and
hence faces the same limitations for long-running, stealthy attacks.

\subsection{APT Models}\label{subsec:related_work-apt_models}
Many models have been proposed to describe multi-stage attacks in general and APT specifically.
Most of them are based on the notion of a chaining of certain attack stages.
Researchers of Lockheed Martin coined the term of an intrusion kill
chain~(IKC)~\cite{model-hutchins-2011}. The model separates an APT into seven distinct phases:
\emph{reconnaissance}, \emph{weaponization}, \emph{delivery}, \emph{exploitation},
\emph{installation}, \emph{command \& control} and \emph{actions on objectives}.
The IKC model is widely accepted and serves as basis for many model-based detection
approaches~\cite{holistic-bhatt-2014, holistic-sexton-2015, holistic-zhang-2017}. Variations of the
IKC exist for different areas of computer science. For example, Hahn et al.\ proposed some
modifications to the IKC, such that it can be adopted to cyber physical
systems~\cite{model-hahn-2015}.

Other attack chain models are based on the analysis of disclosed real-world APT attack reports and
white papers. Following this approach, Siddiqi et al.~\cite{model-siddiqi-2016} and Li et
al.~\cite{model-li-2016} present models with data exfiltration as the overarching goal. In another
analysis Ussath et al.~\cite{model-ussath-2016} find that attackers often use standard tools and
old vulnerabilities (``living off the land'') while they rarely craft sophisticated exploits. Their
attack chain model only considers actions on the victim and does not contain any objectives nor a
reconnaissance phase. All these models are limited to the attacker behavior that was observed in
the analyzed reports. None of these models account for attacker intentions like data corruption or
manipulation of service behavior.

Paul Pols introduced the most comprehensive kill chain-based model, the unified kill
chain~(UKC)~\cite{model-pols-2017}. It extends the IKC model and APT behavior is compared with
red-team intrusions. The UKC model describes attacker behavior in 18 stages and covers host and
network events in advance of the attack, during the attack, and includes campaign objectives. Most
importantly, the UKC model does not enforce all the stages to happen. An APT might not traverse all
possible stages and some stages might be repeated.
The discussed models are not capable of detecting APT attacks by themselves. They only aim to
describe such attacks in the most comprehensive way. The UKC offers the most flexibility without
limiting itself to specific APT attacks to date.

\subsection{APT Detection}\label{subsec:related_work-apt_detection}
APT detection approaches can be divided in two major areas: First, there are approaches that
emphasize a single APT stage and focus on the detection of that particular stage. The existence of
an APT attack is expected when the stage is detected. Second, there are approaches that attempt to
detect and connect multiple stages of an APT model to obtain a more comprehensive view on the
attack.

Bortolameotti et al. proposed a system called DECANTeR~\cite{cnc-bortolameotti-2017} and introduce
the idea of application fingerprinting. A normality model is trained that correlates applications
with their usual HTTP requests. DECANTeR does not require any knowledge about malware and the
generated model is solely based on benign network data. This becomes a drawback when the training
data already contains malware traffic. Furthermore, installing new applications like a different
browser requires to re-train DECANTeR.

Bhatt et al.\ present a framework that incorporates various data sources like IDS or firewall
alerts and host or email logs~\cite{holistic-bhatt-2014}. Their approach uses assumptions on
attacker behavior to map detected attacks to stages in the IKC model.
Similarly, AUSPEX~\cite{holistic-marchetti-2016} incorporates logs from many sources. The authors
focus on the detection of C\&C, lateral movement, and data exfiltration because these stages
exhibit a notable network and host footprint. The framework produces a ranked list of suspicious
hosts to ease the work of analysts.

Generally, the multi-stage contextualization approaches are quite accurate when it comes to the
detection and connection of attack stages. However, they mostly lack the ability to cope with slow
and stealthy attacks. Some approaches have issues to scale along with the amount of monitored data,
in particular those approaches that leverage host monitoring and logs.

\section{Multi-Stage Attack Detection via Kill Chain state machines}\label{sec:system}

\begin{figure}[ht]
    \centering
    \includegraphics[width=\linewidth]{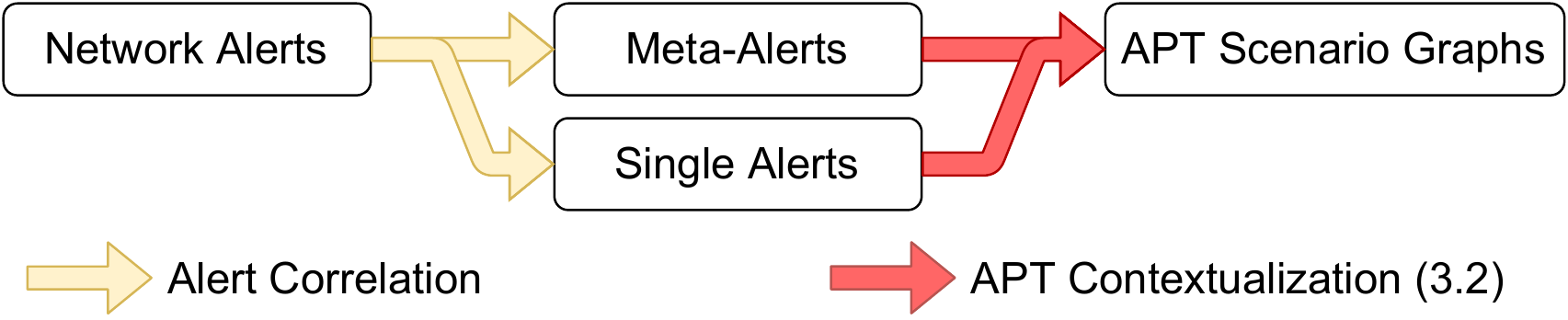}
    \caption{Process overview of multi-stage attack detection via alert correlation and APT contextualization.}\label{fig:approach-overview}
\end{figure}

This section describes our approach for multi-stage attack detection that is summarized
in~\cref{fig:approach-overview}. We first apply alert correlation to preprocess alerts into
clustered meta-alerts and unclustered single alerts. Next, we feed both types of alerts to our APT
detection and contextualization approach. We assign potential attack stages to alerts and link
subsequent stages according to our \emph{kill chain state machine~(KCSM)} to generate \emph{APT
scenario graphs}. These scenarios reveal potential multi-stage atttacks present in the alert data
and aid the human analyst during incident triage and mitigation.
The remainder of this section is structured as follows. In~\cref{subsec:system-state-machine} we
describe two variants of a KCSM that we derived from the unified kill
chain~(UKC)~\cite{model-pols-2017}. Our formalization makes the existing model more actionable and
lays the groundwork for our approach. Next, we describe our algorithm to \emph{characterize and
detect multi-stage attacks} in~\cref{subsec:system-algorithm}. We show how potential attack stages
can be derived from single alerts and meta-alerts and leverage the state machine to connect these
attack stages to expressive \emph{APT scenario graphs}.

\subsection{Kill Chain state machine (KCSM)}\label{subsec:system-state-machine}
A segregation of a network into multiple network zones, which is standard in security-sensitive
organizations, forces an APT actor to follow a kill chain model~\cite{holistic-bhatt-2014}. The
unified kill chain (UKC)~\cite{model-pols-2017} is the most comprehensive model we could find in
related work and was thus chosen as basis for our approach. To make the model actionable for our
algorithms, we formalized it to a state machine that we describe next.
%
\begin{figure*}
    \subfloat[(Full) kill chain state machine (KCSM)\label{fig:ukc-full}]{%
        \includegraphics[width=.55\linewidth]{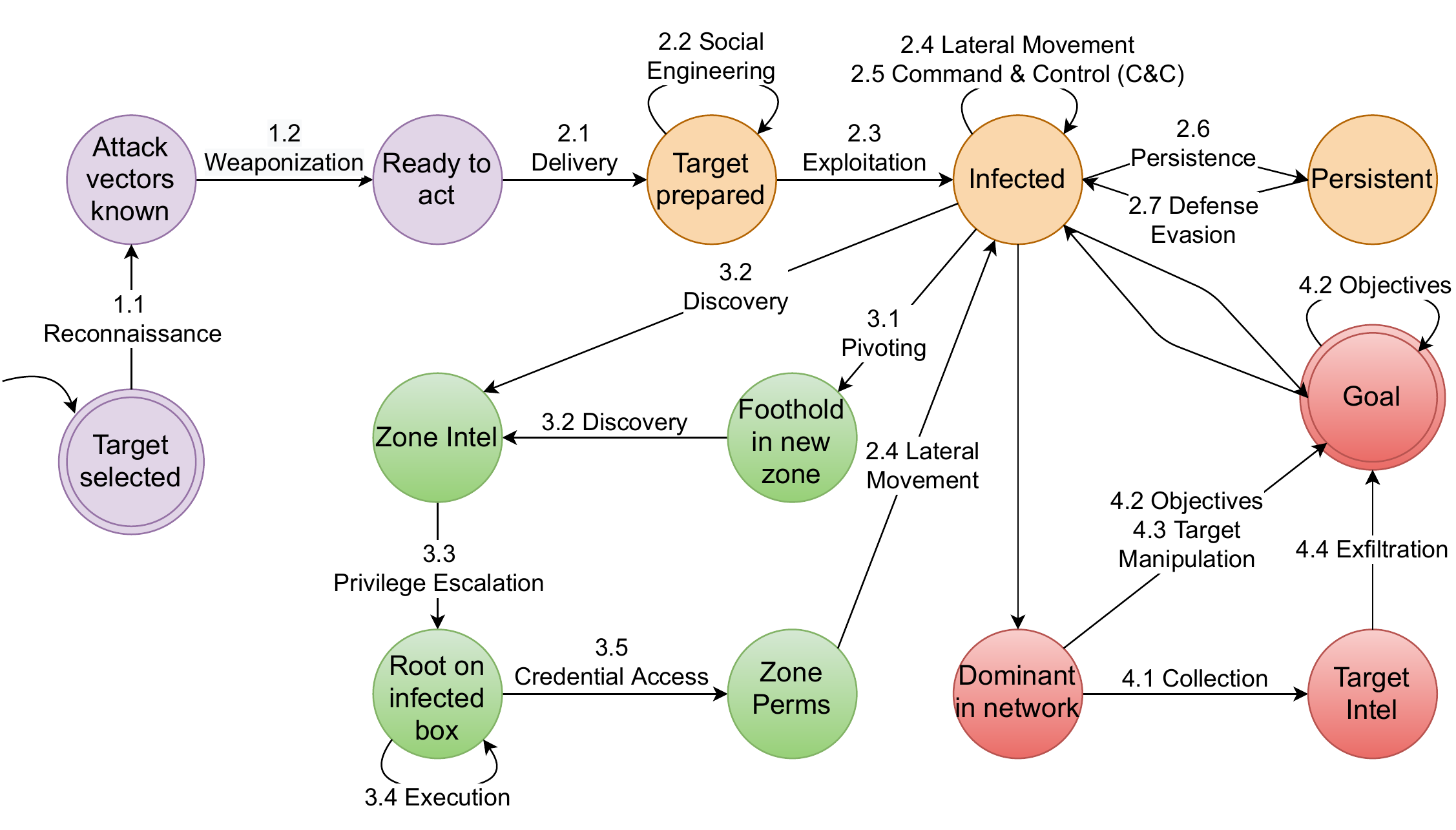}
    }
    \subfloat[(Reduced) network kill chain state machine (NKCSM)\label{fig:ukc-reduced}]{%
        \includegraphics[width=.415\linewidth]{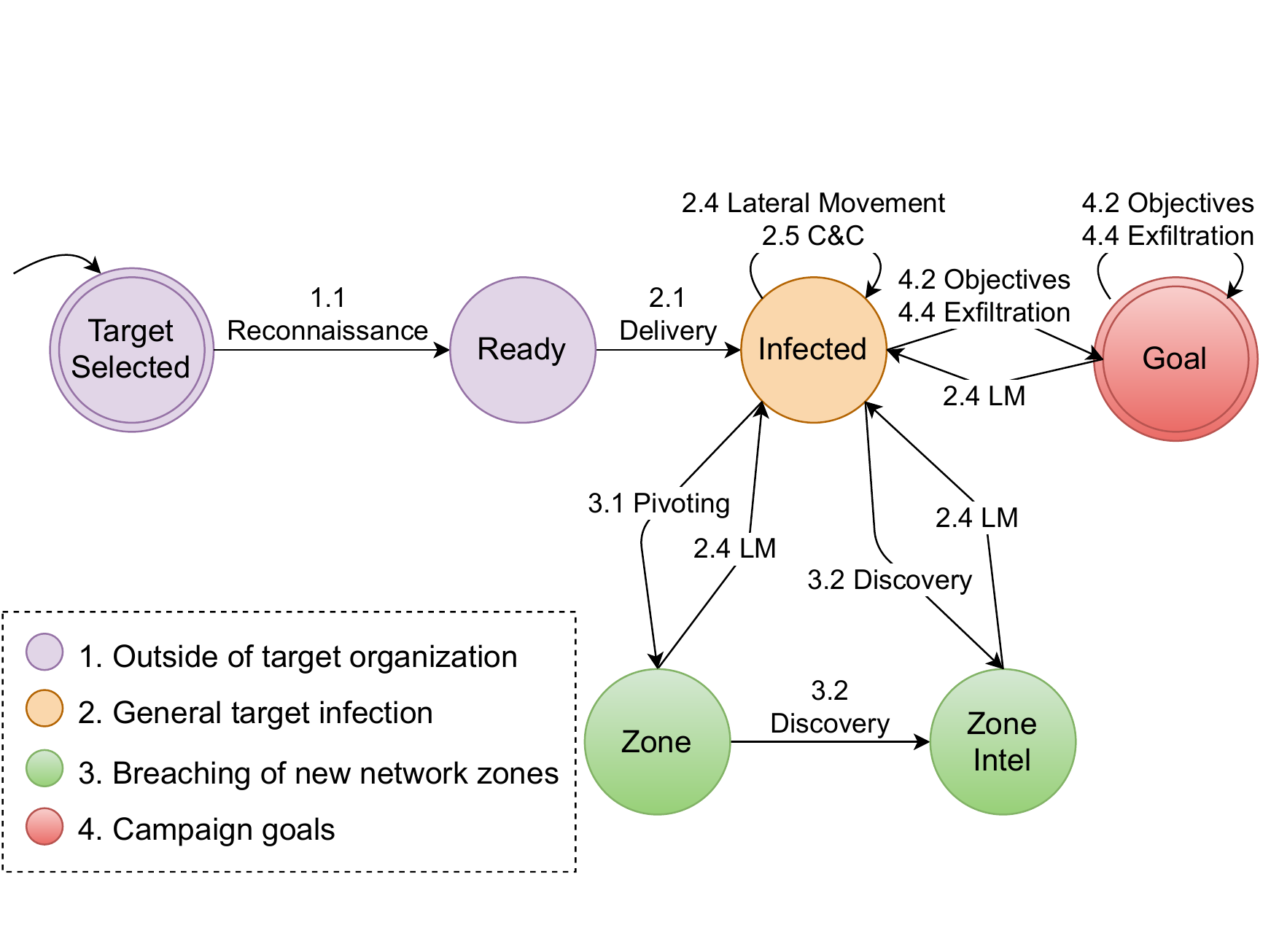}
    }
    \caption{Two APT state machines derived from the 18 stages described in the unified kill chain~(UKC).~\cite{model-pols-2017}}\label{fig:dfa-full}
\end{figure*}
%
The original UKC involves 18 stages (c.f.:~\cite{model-pols-2017}). Naturally, each APT attack is
unique. Some stages might occur repeatedly, others might be left out. Nonetheless, there is a
certain ordering of attack stages that cannot be changed. For example, malware \emph{Delivery}
necessarily comes before \emph{Command \& Control} (C\&C) or \emph{Lateral Movement} (LM).
Likewise, some stages might reoccur. When a network zone is \emph{Pivoted} and new hosts are
infected, C\&C~and LM might be observed repeatedly. We formalize the entirety of connections and
conditions as a finite-state machine in~\cref{fig:ukc-full} that we call the \emph{kill chain state
machine} (KCSM). We map APT stages to transitions as they represent attacker actions. States in
KCSM represent the campaign progress with ``\emph{Target selected}'' as the common start state. As
the goals of APT campaigns can be quite diverse from espionage to manipulation and data theft, the
end state is abstractly labeled ``\emph{Goal}''. Overall, both states and transitions are divided
into four major categories:
\begin{enumerate}
    \item Outside of target organization: States and transitions in this category can usually not be detected or require massive effort as the attackers do not interfere with the target network yet.
    \item General target infection: This category encompasses all states and transitions related to the infection of new hosts in already breached zones. Efficient detection usually requires a mixture of host and network information.
    \item New network zone breaching: States and transitions in this category describe the attacker's progress towards new network zones which are not yet discovered. Detection of actions in this category can largely be achieved via network data analysis as most operations involve movement between hosts.
    \item Campaign goals: This category of states and transitions is highly abstract as it is usually unknown what goals the campaign aims to fulfill. Detection is therefore also difficult with \emph{4.4 Exfiltration} as the only clearly defined transition.
\end{enumerate}

\paragraph*{Outside of target organization (purple)}
 An APT campaign starts when the attacker has selected a target. \emph{1.1 Reconnaissance} follows
 to determine possible attack vectors and to choose an attack strategy. The attack is tailored to
 the target and malware is \emph{1.2 Weaponized} with an exploitable bug.
\paragraph*{General target infection (orange)}
Once the APT actors are ready to act, they \emph{2.1 Deliver} the weapon to the target,\eg{}a
malicious Dropbox implant or a water-holed web server. The next goal is to get the delivered weapon
executed in the target network. Therefore, tactics like \emph{2.2 Social Engineering} might be
employed. After the initial \emph{2.3 Exploitation}, KCSM reaches the central state of infection.
From here on, all further APT actions take place inside the target's infrastructure. The central
state of \enquote{Infected} allows an APT to take various paths. Colors in KCSM states indicate
different action paths that can be pursued. The states for infection and persistence are marked
orange. Whenever a new box is infected, malware may be \emph{2.6 Persisted} and \emph{2.7 Defense
Evasion} techniques are applied. It is possible to take actions in form of \emph{2.5 C\&C} and
\emph{2.4 Lateral Movement (LM)}, just circling in the infected state. LM is considered the
movement to a known target. Identifying new targets, however, falls into the third category.
\paragraph*{Breaching of new network zones (green)}
An APT might need to \emph{3.1 Pivot} from an infected box or might already be in a position where
it is worthy to start \emph{3.2 Discovery}. The \emph{3.2 Discovery} action is similar to \emph{1.1
Reconnaissance} but is conducted in the internal networks of an organization. Thereafter, the zone
is known. Once the attacker \emph{3.3 Escalates Privileges}, they may \emph{3.5 Access Credentials}
to gain network-wide permissions. After that \emph{2.4 LM} is possible within the new network zone,
which brings the attacker back to the general state of infection.
Note that KCSM makes the entire process of network zone breaching an optional path. Controlling the
infection or spreading it to new machines, is always possible via taking the circular \emph{2.4 LM}
and \emph{2.5 C\&C} transitions in the ``Infected'' state. However, this path would presume that
the attacker possess comprehensive knowledge of the network.
\paragraph*{Campaign goals (red)}
The last category of states marks the path for acting towards objectives. From the state of
infection, the attacker can reach a predominant network position or even directly proceed to the
campaign goal. These unlabeled transitions indicate that APT campaign goals can be highly diverse.
It is not required for an APT to \emph{4.1 Collect} or \emph{4.3 Manipulate} data. Similarly,
\emph{4.4 Exfiltration} is not required to happen. However, in case data is exfiltrated, the
attacker is required to collect the targeted data first. The unlabeled backwards transition from
``Goal'' to ``Infected'' state symbolizes how APTs remain in a system as long as possible. Acting
on objectives, exploring new network zones, controlling and spreading are not necessarily bound to
ever finish.
%
It is important to note that we can differentiate the APT stages between stages that
\emph{compromise new hosts} and others that do not. The simplest example for the first category
would be \emph{1.3 Delivery} or \emph{3.1 Pivoting} while \emph{2.5 Command \& Control} or
\emph{4.4 Exfiltration} do not compromise new hosts. We will use this information later in the
contextualization approach to improve the results.
\paragraph*{Network kill chain state machine (NKCSM)}
APT stages can be observed in many parts of a compromised network. \emph{2.1 Delivery} can happen
via the network in the form of an email or offline via an infected thumb drive. \emph{2.3
Exploitation} or \emph{3.3 Privilege Escalation} are solely host-level activities and \emph{3.1
Pivoting} is often both a network and a host action. Monitoring the entire network of an
organization is expensive. But monitoring every individual device of an organization is almost
impossible~\cite{exfil-marchetti-2016}. With a network IDS like \emph{Zeek}~\cite{paxson1999bro} it
is possible to monitor the network traffic even of large networks with little administrative
overhead. In contrast, collecting, shipping, and unifying traces from thousands of heterogeneous
system components is more challenging. Interestingly, about half of the state transitions (and thus
APT stages) might be observable on network level. The other half of all transitions may be observed
on host level or even outside of the organizations scope. To highlight this, we present the
\emph{network kill chain state machine~(NKCSM)} in~\cref{fig:ukc-reduced}, a version of KCSM
reduced to all stages that might be detectable at the network level.
%
NKCSM keeps the start and end states from the original state machine (\emph{Target Selected} and
\emph{Goal}). APT stages that either occur outside of the target's network (\emph{1.2
Weaponization}, \emph{2.1 Social Engineering}) or on host level (\emph{2.3 Exploitation}, \emph{3.3
Privilege Escalation}, \emph{3.4 Execution}, \emph{3.5 Credential Access}, \emph{2.6 Persistence},
\emph{2.7 Defense Evasion}, \emph{4.1 Collection} and \emph{4.3 Target Manipulation}) are excluded.
This reduces the total number of states from thirteen to six, while preserving core semantics of
KCSM. The numbering of APT stages is preserved for simplicity. The general progression from initial
infection over optional lateral movement and zone breaching to action on objectives is still
present in NKCSM. However, all remaining transitions are network-based and thus potentially
observable via a network-based IDS.

\subsection{APT Detection \& Contextualization}\label{subsec:system-algorithm}
The states and transitions formalized in the NKCSM can now be used to trace a multi-stage attack
campaign. However, we still lack a way to map network alerts to potential APT stages. Looking back
at the NKCSM, we quickly realize that all APT stages still present in the state machine are
characterized by movement between specific network zones. Thus, we can use the \emph{network
direction} of an alert to derive potential attack stages it can represent.
We define the network direction of a (meta-)alert to be the transition between the network zones of
the source and destination IP of the alert. A network zones describes one or more subnets of the
same trust level. Examples of typical network zones in an enterprise network are \texttt{external},
\texttt{intranet}, \texttt{datacenter}, and \texttt{dmz}. Given the topology information of the
target network, we can assign the network direction for any network-based alert. This makes our
approach extremely flexible as it does not require special stage information in our source alerts.

\begin{table}
    \centering
    \caption{Network-visible stages of an APT attack. The stage abbreviations are used throughout the
paper.}\label{tb:apt-stages-network}
    \normalsize
    \begin{tabular}{lll}
        \toprule
        Full Name               & Abbr.       & Network direction \\
        \midrule
        Reconnaissance          & \textbf{R}  & \(Z_0 \rightarrow Z_i\) \\
        Delivery                & \textbf{D1} & \(Z_0 \rightarrow Z_i\) \\
        Delivery (2nd stage DL) & \textbf{D2} & \(Z_i \rightarrow Z_0\) \\
        C\&C                    & \textbf{C}  & \(Z_i \rightarrow Z_0\) \\
        Lateral Movement        & \textbf{L}  & \(Z_i \rightarrow Z_j\) \\
        Discovery (``Scan'')    & \textbf{S}  & \(Z_i \rightarrow Z_j\) \\
        Pivoting                & \textbf{P}  & \(Z_i \rightarrow Z_j,~~ i \neq j\) \\
        Exfiltration            & \textbf{E}  & \(Z_i \rightarrow Z_0\) \\
        Objectives              & \textbf{O}  & \(Z_i \rightarrow Z_j\) \\
        \bottomrule
    \end{tabular}
\end{table}

\paragraph*{Attack stage assignment}
\cref{tb:apt-stages-network} shows which network directions are found in each stage of APT attack
stage. As expected, there is an overlap between the stages,\eg{}a meta-alert from \(Z_0\)
(Internet) to \(Z_i\) might be an indicator for either Reconnaissance or Delivery. Thus, our goal
is not to assign a single APT stage to each alert and meta-alert but rather a set of potential
stages. Given the information from~\cref{tb:apt-stages-network}, we obtain four distinct sets of
potential APT stages depending on the network direction of alert. Outgoing connections to the
Internet (\(Z_0\)) are labeled with $[$D2, C, E$]$ and incoming connections with $[$R, D1$]$.
Internal connections are tagged either with $[$L, S, O$]$ if both hosts are part of the same zone
or $[$L, S, P, O$]$ if the zones differ between source and destination host.

\begin{figure*}
    \includegraphics[width=\linewidth]{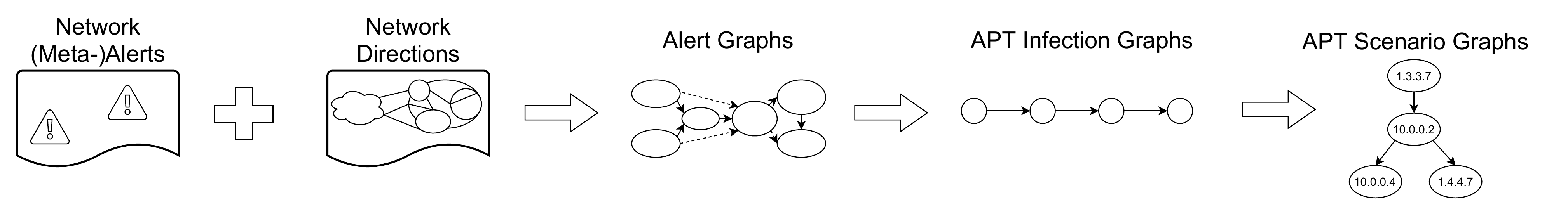}
    \caption{Overview about our approach to APT Contextualization via (N)KCSM.}\label{fig:apt-overview}
\end{figure*}

We have now introduced all fundamentals required for our algorithm. Given the target network
topology, we can assign potential APT attack stages to alerts and meta-alerts that we can then
connect and combine to construct potential multi-stage attack scenarios based on the NKCSM.
\cref{fig:apt-overview} contains all relevant stages of our algorithm. First, we assign potential
stages to alerts and meta-alerts obtained from alert correlation. Next, we build an alert graph by
linking potential predecessor and successor stages based on the NKCSM to obtain \emph{alert
graphs}. These graphs are then reduced and aggregated to \emph{APT infection graphs}. Due to the
construction this graph can still contain multiple potential multi-stage attack campaign, so we
extract multiple distinct \emph{APT scenario graphs}. Finally, the set of scenarios is
\emph{deduplicated} and \emph{pruned} to remove duplicate or trivial scenarios. In the remainder of
the section, we explain all of these steps in more detail.

\paragraph*{Example scenario}
Before we continue to the algorithm, we present a small artificial multi-stage attack scenario,
that we use as an example throughout the paper to illustrate the process. In this scenario, the
target network is divided in two zones \(Z_1\) and \(Z_2\) with the following IP subnets:
%
\begin{align*}
Z_1: & 10.1.0.0/16 \\
Z_2: & 10.2.0.0/16
\end{align*}
The attacker in our scenario aims to find valuable hosts in \(Z_2\) and uses three attacking
machines with public IP addresses (\(4.4.4.4\), \(1.3.3.7\), \(1.4.4.7\)) throughout the campaign.
While a real APT campaign would perform more attack steps, especially after valuable services are
discovered, this example is intentionally kept small. The attacker performs the following five
steps chronologically:

\begin{enumerate}
    \item \emph{Reconnaissance:} The attacker uses an Internet host to scan four hosts in \(Z_0\) (\(4.4.4.4~\rightarrow~[\)\(10.1.0.1\), \(10.1.0.2\), \(10.1.0.3\), \(10.1.0.4]\)).
    \item \emph{Delivery:} The attacker uses another Internet host to deliver a malware-dropper to a host in \(Z_0\) (\(1.3.3.7~\rightarrow~10.1.0.4\)).
    \item \emph{Delivery (Download):} The infected machine downloads a second-stage malware from the Internet (\(10.1.0.4~\rightarrow~1.4.4.7\)).
    \item \emph{Pivot:} The malware pivots the infection to two hosts in \(Z_1\) (\(10.1.0.4~\rightarrow~[10.2.0.1,~10.2.0.3]\)).
    \item \emph{Discovery:} The malware scans the three remaining machines in \(Z_1\) for valuable services (\(10.2.0.3~\rightarrow~[\)\(10.2.0.2\), \(10.2.0.4\), \(10.2.0.5]\))
\end{enumerate}

\begin{figure*}
    \centering
    \includegraphics[width=.80\linewidth]{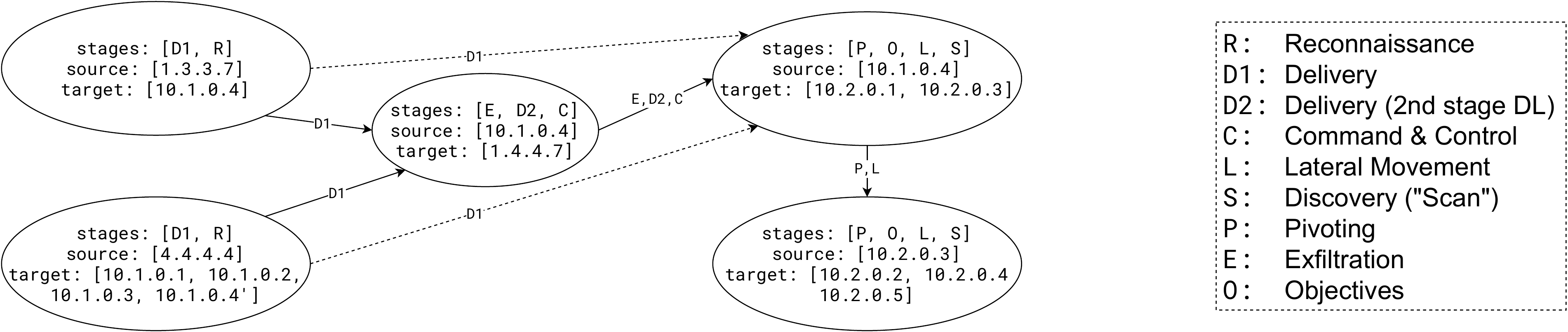}
    \caption{\emph{Alert graph} comprising six alerts generated from the example scenario.}\label{fig:meta-alert-graph}
\end{figure*}

\paragraph*{Alert graph}
With the implied transition order from NKCSM and the mapping of alerts to potential APT stages, we
can build a directed graph based on pre- and postconditions of each state. Each meta-alert results
in a node with the potential stages, sources and targets of the attack attached as additional
labels. Two nodes \(u,v\) are connected with an edge \(e\) if three conditions are met:
\begin{enumerate}
    \item The \emph{latest} timestamp of \(u\) is smaller than the \emph{earliest} timestamp of \(v\).
    \item The potential APT stages of \(u\) contain at least one stage, that is a precondition for any APT stage of \(v\). If multiple stages match, the ones that \emph{compromise new hosts} are preferred.
    \item The \emph{source} and \emph{target} IP addresses of \(u\) and \(v\) match according to the APT stages of \(u\). For stages that \emph{move infection} (D1, L, P) the \emph{target} address of \(u\) matches a \emph{source} address of \(v\). Other APT stages, that do not compromise new hosts, are valid if any \emph{source} IP address overlaps between \(u\) and \(v\).
\end{enumerate}
The resulting edge \(e\) is then labeled with the set of APT stage preconditions,\ie{}the set of
APT stages of \(u\) that represent a precondition to any stage of \(v\).
\cref{fig:meta-alert-graph} shows such a alert graph that contains six alerts derived from the five
stages in our example. For demonstrative purposes we assume that an IDS produced a alert for each
malicious action without any false positives.

The alert graph already shows promising results. The path the attackers took throughout the
zones is clearly visible. However, we can already see at least two suboptimal properties of the
graph: First, the APT stage labels on the edges are broad and thus sometimes contain more APT
stages than only the correct one. This is not a problem, as the more relevant information,\ie{}IP
addresses and alert identifiers, is contained in the nodes anyway. Second, the alert graph
contains edges that do not add any useful information (indicated by dashed lines in the figure) as
they lie on a path shorter than the longest path between two nodes. This means, they discard the
additional information that is present on the nodes that are not included on these paths. While it
is possible, that this represents the actual attack and the skipped node is the result of a false
positive alert, the longest path offers more information to the threat hunter (namely the hosts to
investigate for further indicators of compromise). Additionally, this type of graph can grow quite
large quite easily. Due to the loose requirements for linking two nodes, resulting graphs can be
almost fully connected in real-world scenarios. This would not efficiently support human
threat-hunters as they cannot extract meaningful information anymore. However, our example graph
is relatively compact, it only represents a small attack without any false positives.

\paragraph*{APT infection graph}
We can address the problems of the alert graph by \emph{reducing} the graph density through
\emph{elimination} of obsolete information and \emph{aggregation} of existing information.
We start the graph consolidation on the node that does not possess any outgoing edges. We can
guarantee the existence of at least one such node due to our timestamp requirement on the edges
(condition 1 for connecting two nodes in the alert graph). Starting from this node, we
recursively iterate through the incoming edges while aggregating those with identical APT stage
labels. Longest paths are preferred during the iteration and paths shorter than the longest path
between two nodes are discarded. The \emph{source} and \emph{target} sets of the connected nodes
are combined for matching edges to obtain the set union.

\begin{figure*}
    \centering
    \includegraphics[width=.85\linewidth]{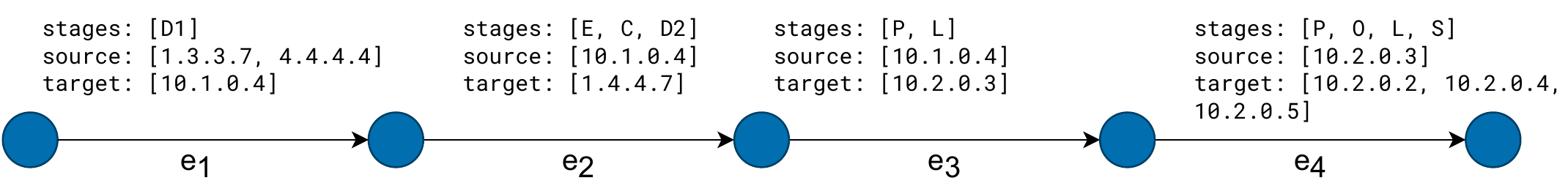}
    \caption{\emph{APT infection graph} for the example scenario without false positives.}\label{fig:infection-graph}
\end{figure*}

\begin{figure}
    \centering
    \includegraphics[width=\linewidth]{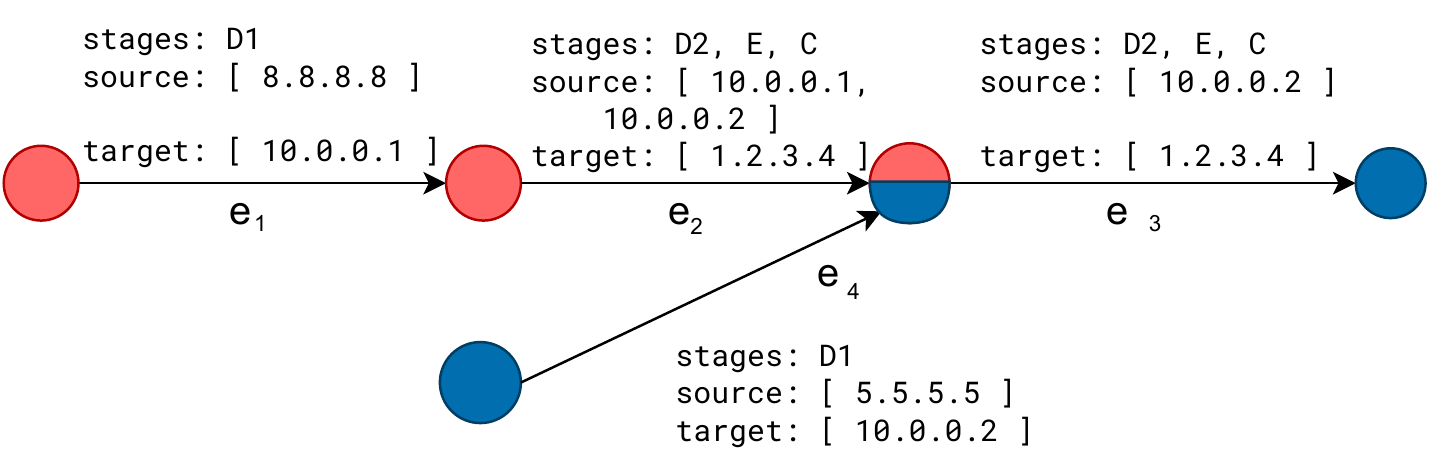}
    \caption{Example for a transitively invalid \emph{APT infection graph}.}\label{fig:infection-graph-invalid}
\end{figure}

The compact graph obtained by this process is called \emph{APT infection graph}. In this graph,
nodes mark APT campaign progress, while the edges represent the APT stages with the related
information such as involved IP addresses and alert identifier. \cref{fig:infection-graph} shows
the APT infection graph for our example scenario.
It is significantly smaller than before, while retaining the important information about IP
addresses and stages. The progress of the potential APT campaign is clearly visible, and the
reduced branching helps focusing on the important information for threat hunters. The combination
of the set of potential APT stages and IP addresses directly hints at how the hosts should be
further investigated as the different stages usually leave distinct indicators of compromise on the
machine.
When we described KCSM and NKCSM, we differentiated between stages that compromise new hosts and
stages that do not. Until now we only used this information to prioritize stages while constructing
alert graphs. As a result, the APT infection graph might contain edge pairs that represent
consecutive stages in the state machine (and thus are valid in the current graph) but have a
non-infected host as the source for the second stage. An example for that is given
in~\cref{fig:infection-graph-invalid}.

%
In the figure we see that host $10.0.0.1$ gets infected from the Internet ($e_1$). This is followed
by some outbound connections from the hosts $10.0.0.1$ and $10.0.0.2$ ($e_2$) and another outbound
connection from $10.0.0.1$ only ($e_3$). While the consecutive stages D1 $\rightarrow$ [D2, E, C]
$\rightarrow$ [D2, E, C] are valid, the host $10.0.0.2$ was never infected on the red path ($e_1,
e_2$) and thus cannot be responsible for the outbound connection in $e_3$. If we follow the blue
path ($e_4, e_3$) the host is indeed infected, but the source IP for the preceding \emph{Delivery
stage} differs ($5.5.5.5$). If we consider this transitive validity, the graph actually contains
two distinct potential APT scenarios that partially overlap in hosts and the correct solution would
be to extract these two scenarios respectively.

\begin{figure}
    \centering
    \includegraphics[width=.85\linewidth]{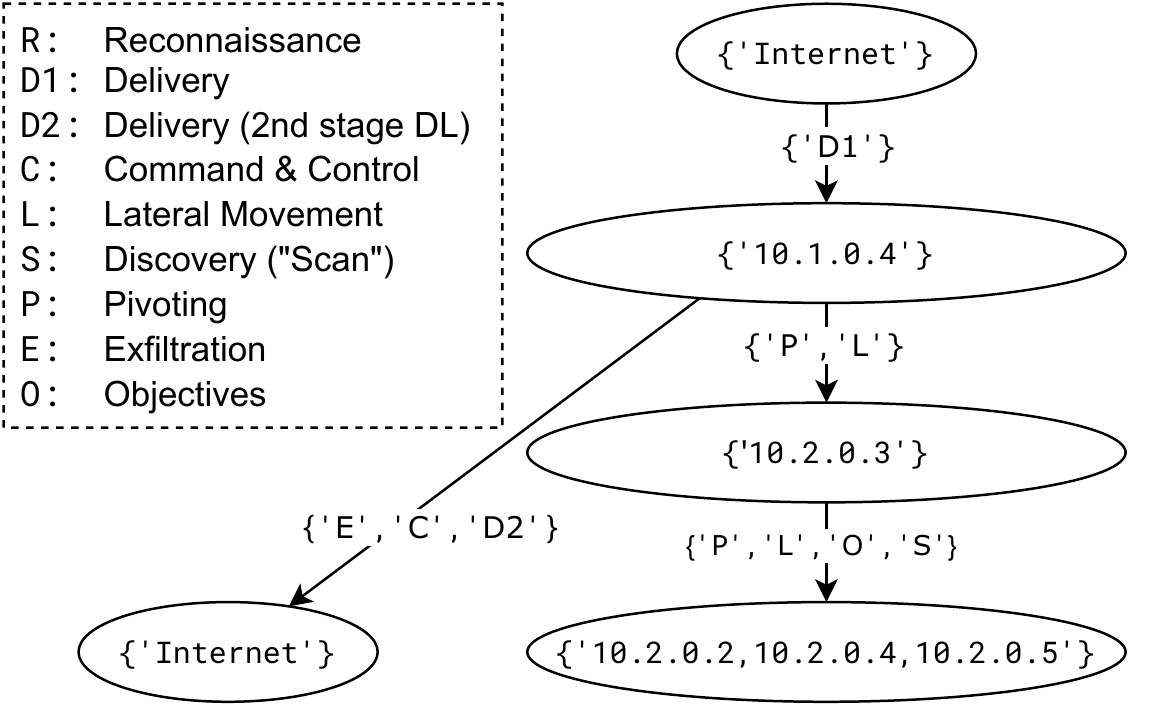}
    \caption{\emph{APT scenario graph} representing the APT campaign from the example scenario.}\label{fig:scenario-graph}
\end{figure}

\paragraph*{APT scenario graph}
Overall, \emph{APT infection graphs} provide a decent overview about potential multi-stage attack
campaigns. However, as we mentioned they can still contain edges that can be considered invalid
when compared with our original state machine NKCSM. Additionally, they might contain multiple APT
scenarios as separate paths. To address this, we introduce another optimization step to obtain the
final graph representation for potential APT scenarios.
%
\emph{APT scenario graphs} can be obtained by extracting transitively valid paths from APT
infection graphs. The structure closely resembles the KCSM. Edges are labeled with sets of
potential APT stages while nodes contain the involved IP addresses in the respective attack steps.
\cref{fig:scenario-graph} shows the APT Scenario graph that the algorithm produces for our example
scenario. The graph is concise and supports the human analyst in combating the \emph{alert fatigue}
by highlighting the affected internal hosts and the progress of the potential APT campaign.
External IP addresses are replaced with the \emph{'Internet'} label to limit the potential state
expansion. These are of secondary interest to the analyst anyway, as they can only directly
investigate internal hosts. However, they are not \enquote{lost} as the full alert data is
available through the alert identifiers attached to the graph (left out here for brevity). To sum
up, the APT scenario graph not only represents a \emph{significant reduction} from the original
alert set it was generated of but also offers essential context to human threat hunters through the
campaign progress.

The final set of \emph{APT scenario graphs} can be further reduced and optimized. Due to the
process for graph construction, we can sometimes obtain two graphs, where one is an
\emph{isomorphic subgraph} of the other. In our context, this means, that both graphs describe the
same potential APT scenario and the larger graph just contains additional steps that are not
present in the smaller one. Similar to the edge elimination step from \emph{alert graphs} to
\emph{APT infection graph}, we can eliminate these smaller graphs entirely without losing relevant
information. While the problem of subgraph isomorphism is known to be NP-complete, it can often be
solved efficiently. Our prototype uses the implementation from the Python library \texttt{networkx}
that is based on Cordella's work~\cite{cordella-2004} to prune and deduplicate the final result
set. While our prototype does not perform any additional optimization, a real-world deployment can
add additional postprocessing of the final set of APT scenario graphs,\eg{} prioritization of
scenarios that contain a specific critical host or scenarios with the longest chains. As APT and
other multi-stage attacks are highly dynamic and tailored to the target, we want to emphasize the
opportunities for further optimization here, but leave the actual implementation to the respective
target organization.

\subsection{Summary}\label{subsec:system-summary}
In this section, we described the two core contributions of our paper. We first introduced our
formalization of the unified kill chain~\cite{model-pols-2017}, the \emph{kill chain state
machine~(KCSM)} and a reduced variant specific to network-visible stages (\emph{network kill chain
state machine~(NKCSM)}). Second, we detailed our graph-based algorithm to produce \emph{APT scenario
graphs} from a set of network alerts and meta-alerts and the topology information of the target
network. The scenarios not only represent a significant reduction when compared to the total alert
volume, but also offer additional context to human analysts during incident triage and further
investigations of potential multi-stage attack present in the network.

\section{Evaluation}\label{sec:eval}
We evaluate our algorithm for APT Contextualization in two distinct ways. First, we apply our
complete approach, from alert generation to APT contextualization, to the entire unmodified
CSE-CIC-IDS2018~\cite{ids-sharafaldin-2018} dataset. While this dataset does not contain any APT
activity by itself, this experiment demonstrates the significant alert volume reduction we can
achieve in a realistic scenario. Furthermore, we highlight the real-world applicability of our
approach as we process raw network traffic instead of commonly used alert data.
Second, we designed an APT attack campaign based on real CVEs, injected PCAPs of the attacks into
CSE-CIC-IDS2018 and applied our pipeline again. We explain how the resulting \emph{APT Scenario
Graphs} provide unique insights into the APT campaign and show how the additional context can aid
mitigation measures.
This section begins with a brief overview of both the original CSE-CIC-IDS2018 dataset as well as
the modified one including our custom APT campaign. Next, we summarize the experimental setup
including the different tools and algorithms used for alert generation and correlation. In the last
part, we discuss the results obtained on both the unmodified dataset as well the injected APT
campaign and show how our approach achieves a significant volume reduction of the alert set and
provides additional context about APT campaigns to threat hunters.

\subsection{Data Sets}
\begin{table}
    \centering
    \caption{CSE-CIC-IDS2018: Overview}\label{tb:eval-ids2018}
    \begin{tabular}{ll}
        \toprule
        Property               & Value \\
        \midrule
        \# Subnets/Zones       & 6 + \emph{Internet} \\
        \# Target Hosts        & 450                 \\
        \# Attacker Hosts      & 50                  \\
        \# Connections         & \IMconn{}           \\
        \# (unrelated) attacks & 7                   \\
        Duration               & 10 days             \\
        Size in GB             & \IMsize{}           \\
        \bottomrule
    \end{tabular}
\end{table}
To the best of our knowledge, there are no public network traffic data sets containing real APT
campaigns. In combination with the highly dynamic nature of APT campaigns this renders the
evaluation of detection and correlation approaches difficult. We address this problem by taking a
well-known data set for intrusion detection, namely CSE-CIC-IDS2018~\cite{ids-sharafaldin-2018},
and carefully inject a hand-crafted APT campaign into it. To model the individual attack steps of
the campaign, we chose known vulnerabilities that were actively exploited.
The CSE-CIC-IDS2018 data set is intended for the evaluation of network IDSs and contains seven
distinct attacks spanning ten days across six internal network zones/subnets as well as benign
traffic. \cref{tb:eval-ids2018} shows some key characteristics of the data set. The scope of six
internal zones as well as 450 hosts should match a small to mid-sized company well. The duration of
ten days is quite small when considering APT scenarios, however it still allows for a multi-step
attack campaign with certain delays in between. Although no APT campaign is present in the original
data set, some of the seven contained attacks resemble single steps of an APT campaign,\eg{}an
infiltration attack. Additionally, some other attacks, such as DDoS or password guessing via
Bruteforce usually produce large amounts of traffic and a large number of alerts unrelated to the
APT campaign.

To evaluate the detection performance of our contextualization approach, we designed a realistic
APT campaign. As the source data set contains network traffic over ten days, we could not exceed
this duration in our scenarios. We named this campaign IDS2018-APT and its attack steps are
summarized in~\cref{tb:eval-apt-min-attacks}. The APT campaign proceeds as follows:
%
\begin{table}
    \centering
    \caption{IDS2018-APT: Campaign overview}\label{tb:eval-apt-min-attacks}
    \begin{tabular}{clrr}
        \toprule
        Day  & Attack                      & Source       & Target       \\
        \midrule
        1    & \emph{EternalRomance} RCE   & 1.1.13.37    & 172.31.64.67 \\
        1    & 2nd stage trojan download   & 172.31.64.67 & 12.34.12.34  \\
        4    & Cosmic Duke C\&C            & 172.31.64.67 & 1.1.14.47    \\
        8    & PS-EXEC via SMB             & 172.31.64.67 & 172.31.69.20 \\
        10   & Data exfiltration via HTTPS & 172.31.69.20 & 1.1.15.57    \\
        \bottomrule
    \end{tabular}
\end{table}


\begin{enumerate}
    \item On day 1 the attackers perform a Remote Code Execution (RCE) via the \emph{EternalRomance} exploit on a vulnerable host in the R\&D department of the target organization with the IP address \(172.31.64.67\).
    \item Directly afterwards, the malware downloads a second stage trojan from a compromised Internet host with the IP address \(12.34.12.34\) and persists on the machine.
    \item Three days later, the malware performs Command \& Control (C\&C) communication to the attacker-controlled Internet host \(1.1.14.47\).
    \item On day 5 the malware moves laterally to a server machine \(172.31.69.20\) via \texttt{PS-EXEC}, a tool for legitimate Windows remote administration.
    \item On day 10 the compromised server exfiltrates a copy of a core database to the attacker-controlled Internet host \(1.1.15.57\).
\end{enumerate}
Overall, IDS2018-APT involves four malicious hosts from the Internet (\(1.1.13.37\),
\(12.34.12.34\), \(1.1.14.47\), \(1.1.15.57\)) and two internal hosts that are targeted by attacks
(\(172.31.64.67\), \(172.31.69.28\)). The movement across two zones resembles a potential APT
campaign in which the attackers had inside knowledge about the organization's network
segmentation and directly moved to the target server.

The APT campaign was injected into the CSE-CIC-IDS2018 data set, which contains benign traffic as
well as some unrelated attacks. The PCAP data for the attack steps was collected from various
sources across the web:
\emph{ericconrad.com}\footnote{\url{https://www.ericconrad.com/2017/04/shadowbrokers-pcaps-etc.html}}
for the \emph{EternalRomance} RCE and trojan download, University of
Twente\footnote{\url{https://www.utwente.nl/en/eemcs/scs/downloads/20171127_DEM/}} for the
\emph{Data Exfiltration Malware} samples and more specifically \emph{Cosmic Duke} C\&C traffic and
\emph{github.com/401trg}\footnote{\url{https://github.com/401trg/detections/tree/master/pcaps}} for
PCAPs containing PS-EXEC via SMB. The exfiltration was performed locally, and the traffic recorded
via \texttt{tcpdump}. As mentioned before all attacks were chosen on the basis of previous
exploitation in the wild,\eg{}PS-EXEC is the de facto standard for lateral movement in typical
Windows-based enterprise networks at the time of writing. The various PCAP files were then
carefully rewritten in both IP addresses and timestamps via \texttt{tcprewrite} and
\texttt{editcap} to match our APT campaign steps. This process resulted in a coherent attack
scenario that took place at the time of the original CSE-CIC-IDS2018 data set (02.18--03.18).

\subsection{Experimental Setup}
\begin{figure}
    \centering
    \includegraphics[width=\linewidth]{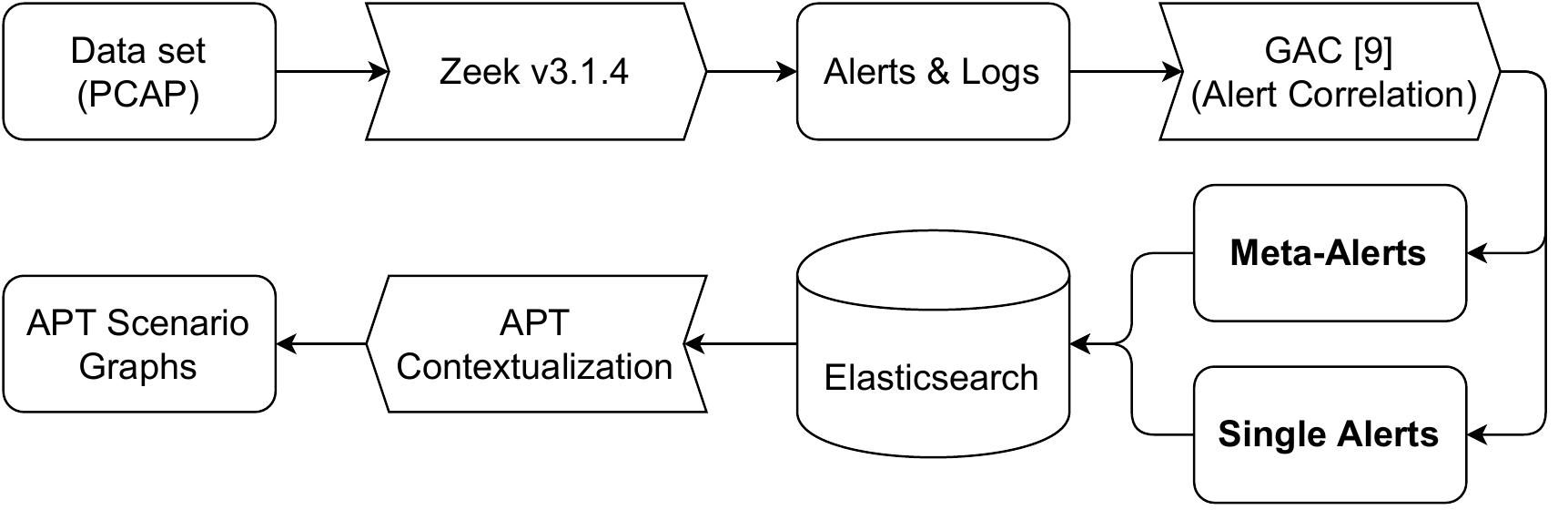}
    \caption{Full experimental setup for End-to-End APT Contextualization.}\label{fig:eval-apt-setup}
\end{figure}
%
\cref{fig:eval-apt-setup} gives an overview across the experimental setup used throughout the
evaluation. The scenario PCAPs were processed by \emph{Zeek} as network IDS, resulting in alerts
and log files. Next, we used Graph-based Alert Correlation (GAC)~\cite{ac-haas-2018} as an
established alert correlation algorithm. This resembles a real-world deployment with Zeek as the
IDS and some alert correlation to reduce alert volume. The processing with GAC yields
\emph{clustered meta-alerts} and \emph{unclustered single alerts}. Both sets of alerts are
persisted in Elasticsearch for later use in the APT contextualization. Our prototype implementation
of the APT contextualization is written in Python and leverages the \emph{networkx} library for
graph processing.

For alert generation we use Zeek v\(3.1.4\) in two distinct configurations that we call MIN and
FULL. MIN loads almost all scripts that Zeek provides per default. Notable exceptions include file
extraction, as the transferred files are not needed in the process, and SSL/TLS/OCSP verification,
as these scripts produce false positives for the self-signed certificates that are used in
CSE-CIC-IDS2018. In addition to the default scripts, we wrote some Zeek scripts tailored to our
organization scenario including detection of downloaded Windows executables and large outgoing
network communication. These organization-specific scripts are \emph{not} written to exactly detect
the attack steps of our APT campaign directly, but rather match the overall scenario of a mostly
Windows-based enterprise network. As a result, they produce false positives especially as the
traffic unrelated to our APT campaign also consists of the same protocols that the scripts
monitor,\eg{} SMB. MIN therefore resembles an organization that customizes Zeek to some extent,
however, does not include any third-party scripts for improved visibility.
The FULL configuration loads all scripts from MIN and adds two well-known third-party scripts,
namely \texttt{mitre-attack/bzar}\footnote{\url{https://github.com/mitre-attack/bzar}} for the
detection of adversarial activity related to Mitre's ATT\&CK framework and
\texttt{0xl3x1/zeek-EternalSafety}\footnote{\url{https://github.com/0xl3x1/zeek-EternalSafety}} for
detecting potentially malicious SMBv1 protocol violations that are used in the \textit{Eternal*}
family of Windows exploits. FULL therefore resembles a realistic setup in high-security
environments, in which security administrators perform proactive threat hunting to detect APT
campaigns. The MITRE ATT\&CK framework is well known in the community and aimed at detecting attack
steps of multi-step attacks such as APT campaigns.
Both configurations are examples for organizations with different threat profiles. MIN includes
less scripts and might therefore miss important attack steps. FULL includes scripts that will
result in an increased alert volume with more false positives, which will also complicate
successful detection of APT campaigns. We include both configurations to show the range of
scenarios our approach addresses.

GAC~\cite{ac-haas-2018} was chosen as state-of-the-art approach for alert correlation. This step is
essential to reduce the alert volume. GAC achieves this via clustering, however any correlation
algorithm can be used as both meta-alerts and single alerts are ingested for the APT
contextualization. GAC was run on batches of all alerts once per day in the data set to simulate a
realistic scenario where batch-based alert correlation algorithms are run in fixed intervals. We
also experimented with DECANTeR~\cite{cnc-bortolameotti-2017} to reveal C\&C communication and
produce stage-alerts accordingly. However, in our experiments, DECANTeR was unable to correctly
identify the C\&C communication in our APT campaign and was thus not included in the final
evaluation. Nonetheless, our approach is not strongly tied to any alert correlation or APT stage
detection scheme. Other alert correlation approaches can be used and stage-specific alerts for any
APT stage are supported.

\subsection{Results}
\begin{table}
    \centering
    \caption{IDS2018: Results Overview}\label{tb:eval-overview-ids2018}
    \begin{tabular}{clrr}
        \toprule
        Level            & Metric                       & MIN                  & FULL \\
        \midrule
        Zeek             & Alerts                       & \IMalerts            & \IFalerts \\
        \midrule
        \mrotrow{4}{GAC} & Meta-Alerts                  & \IMgacalerts         & \IFgacalerts \\
                         & Singleton Alerts             & \IMweakalerts        & \IFweakalerts \\
                         \cmidrule(r){2-4}
                         & Alerts for Contextualization & \IMhighlevelalerts   & \IFhighlevelalerts \\
        \midrule
        \mrotrow{4}{APT} & APT infection graphs         & \IMinfectiongraphs   & \IFinfectiongraphs \\
                         & Total APT scenario graphs    & \IMtotalscenarios    & \IFtotalscenarios \\
                         & Distinct APT scenario graphs & \IMdistinctscenarios & \IFdistinctscenarios \\
                         & Volume reduction             & \IMreductionpc\%     & \IFreductionpc\% \\
        \bottomrule
    \end{tabular}
\end{table}

\subsubsection*{Volume reduction on unmodified data set}\label{subsubsec:results-volume}
In the first experiment, we evaluate the real-world feasibility of our approach by processing the
unmodified CSE-CIC-IDS2018 data set~\cite{ids-sharafaldin-2018}. As the data set does not contain
an APT campaign, all results can be considered false positives. However, the results and especially
the number of \emph{APT scenario graphs} should help to estimate the amount of manual analysis work
that is required to be performed by security personnel when using our approach in day-to-day work.
Thus, the experiment will establish a baseline to compare the following experiment with the
injected IDS2018-APT scenario.
As mentioned before, CSE-CIC-IDS2018 does contain both external and internal attacks unrelated to
the APT campaign,\eg{}brute-forcing, an infiltration attack originating from an internal host or
DDoS originating from a botnet. This should be representative of real-world networks that may be
targeted by outside attacks as well as some internal noise.

\cref{tb:eval-overview-apt} contains the results for both Zeek configurations for the unmodified
data set. The MIN configuration produced \IMalerts\ Zeek alerts that GAC further clustered to
\IMgacalerts\ meta-alerts and \IMweakalerts\ unclustered alerts. Thus \IMhighlevelalerts\ total
alerts were used as input for the APT contextualization. This resulted in \IMinfectiongraphs\
\emph{APT infection graphs} which were split into \IMtotalscenarios\ total \emph{APT scenario
graphs} that were deduplicated and pruned to \IMdistinctscenarios\ graphs. Over the experiment
duration of ten days, this amounts to approximately \IMscenariosperday\ potential APT scenarios per
day that need to be manually analyzed by human analysts. While this number initially may seem high,
it is not unreasonable to assume that a security operation center (SOC) of an organization with 450
networked hosts would be able to accomplish. This is especially likely as the alternative would be
to manually investigate all \IMalerts\ alerts in isolation and without additional context offered
by our APT scenario graphs. Overall our approach yields a \emph{significant volume reduction} to
\textbf{\IMreductionpc\%} of the original alert set.
The results for the FULL configuration are similar and partially even better. In total Zeek
produced \IFalerts\ alerts. This significant increase, when compared to the MIN configuration, is
the result of adding including scripts with higher volume output. By applying GAC, we obtained
\IFgacalerts\ meta-alerts and \IFweakalerts\ unclustered alerts. The APT contextualization of these
\IFhighlevelalerts\ total meta-alerts yields \IFinfectiongraphs\ APT infection graphs and
\IFtotalscenarios\ total APT scenario graphs, respectively. For the FULL configuration, the
deduplication and pruning proved to be highly effective. It reduced the number of potential
scenarios from \IFtotalscenarios\ to \IFdistinctscenarios. When compared to our original alert set
of \IFalerts\ alerts, this number resembles an even \emph{larger volume reduction} to
\textbf{\IFreductionpc\%}. This shows the potential of our approach especially for larger alert
sets.

\begin{table}
    \centering
    \caption{IDS2018-APT: Result Overview}\label{tb:eval-overview-apt}
    \begin{tabular}{clrr}
        \toprule
        Level            & Metric                       & MIN                  & FULL \\
        \midrule
        Zeek             & Alerts                       & \AMalerts            & \AFalerts            \\
        \midrule
        \mrotrow{4}{GAC} & Meta-Alerts                  & \AMgacalerts         & \AFgacalerts         \\
                         & Singleton Alerts             & \AMweakalerts        & \AFweakalerts        \\
                         \cmidrule(r){2-4}
                         & Alerts for Contextualization & \AMhighlevelalerts   & \AFhighlevelalerts   \\
        \midrule
        \mrotrow{4}{APT} & APT infection graphs         & \AMinfectiongraphs   & \AFinfectiongraphs   \\
                         & Total APT scenario graphs    & \AMtotalscenarios    & \AFtotalscenarios    \\
                         & Distinct APT scenario graphs & \AMdistinctscenarios & \AFdistinctscenarios \\
                         & Volume reduction             & \AMreductionpc\%     & \AFreductionpc\%     \\
        \bottomrule
    \end{tabular}
\end{table}

\begin{table*}
    \caption{IDS2018-APT: Groundtruth in Zeek Alerts}\label{tb:eval-apt-zeek}
    \begin{adjustbox}{width=.9\linewidth,center}
    \begin{tabular}[b]{clrrrrrr}
    \toprule
           &                                         & \multicolumn{3}{c}{IDS2018-APT-MIN}                      & \multicolumn{3}{c}{IDS2018-APT-FULL} \\
                                                     \cmidrule(r){3-5}                                          \cmidrule(r){6-8}
    Source & Alert Type                              & \# Alerts       & APT related       & Ratio              & \# Alerts       & APT related       & Ratio \\
    \midrule
    \mrotrow{2}{Zeek}  & Conn::Retransmission\_Inconsistency & \AMconnre & \AMconnretp     & \AMconnretpr       & \AFconnre       & \AFconnretp       & \AFconnretpr       \\
     & SSL::Weak\_Key                                & \AMweakkey      & \AMweakkeytp      & \AMweakkeytpr      & \AFweakkey      & \AFweakkeytp      & \AFweakkeytpr      \\
    \addlinespace
    \mrotrow{10}{Organization-specific} & Org::Stalled\_HTTP\_Connection & \AMstalledhttp & \AMstalledhttptp & \AMstalledhttptpr & \AFstalledhttp & \AFstalledhttptp & \AFstalledhttptpr  \\
     & Org::HTTP\_Windows\_Executable\_Dl            & \AMsighttp      & \AMsighttptp      & \AMsighttptpr      & \AFsighttp      & \AFsighttptp      & \AFsighttptpr      \\
     & Org::NON\_HTTP\_Windows\_Executable\_Dl       & \AMsignonhttp   & \AMsignonhttptp   & \AMsignonhttptpr   & \AFsignonhttp   & \AFsignonhttptp   & \AFsignonhttptpr   \\
     & Org::SMB\_Executable\_File\_Transfer          & \AMsmbtransfer  & \AMsmbtransfertp  & \AMsmbtransfertpr  & \AFsmbtransfer  & \AFsmbtransfertp  & \AFsmbtransfertpr  \\
     & Org::Javascript\_Web\_Injection\_URI          & \AMjswebinj     & \AMjswebinjtp     & \AMjswebinjtpr     & \AFjswebinj     & \AFjswebinjtp     & \AFjswebinjtpr     \\
     & Org::SQL\_Web\_Injection\_URI                 & \AMsqlinj       & \AMsqlinjtp       & \AMsqlinjtpr       & \AFsqlinj       & \AFsqlinjtp       & \AFsqlinjtpr       \\
     & Org::Web\_Login\_Guessing                     & \AMweblogin     & \AMweblogintp     & \AMweblogintpr     & \AFweblogin     & \AFweblogintp     & \AFweblogintpr     \\
     & Org::Large\_Outgoing\_Tx                      & \AMlargetx      & \AMlargetxtp      & \AMlargetxtpr      & \AFlargetx      & \AFlargetxtp      & \AFlargetxtpr      \\
     & Org::Multiple\_Large\_Outgoing\_Tx            & \AMmultilargetx & \AMmultilargetxtp & \AMmultilargetxtpr & \AFmultilargetx & \AFmultilargetxtp & \AFmultilargetxtpr \\
     & Org::Very\_Large\_Outgoing\_Tx                & \AMverylargetx  & \AMverylargetxtp  & \AMverylargetxtpr  & \AFverylargetx  & \AFverylargetxtp  & \AFverylargetxtpr  \\
    \addlinespace
    \mrotrow{3}{BZAR} & ATTACK::Execution            & \AMatkexec      & \AMatkexectp      & \AMatkexectpr      & \AFatkexec      & \AFatkexectp      & \AFatkexectpr      \\
     & ATTACK::Lateral\_Movement                     & \AMatklm        & \AMatklmtp        & \AMatklmtpr        & \AFatklm        & \AFatklmtp        & \AFatklmtpr        \\
     & ATTACK::Lateral\_Movement\_Extracted\_File    & \AMatklmextract & \AMatklmextracttp & \AMatklmextracttpr & \AFatklmextract & \AFatklmextracttp & \AFatklmextracttpr \\
    \addlinespace
    \mrotrow{7}{EternalSynergy} & EternalSafety::DoublePulsar & \AMdouplepulsar & \AMdouplepulsartp & \AMdouplepulsartpr & \AFdouplepulsar & \AFdouplepulsartp & \AFdouplepulsartpr \\
     & EternalSafety::EternalBlue                    & \AMetblue       & \AMetbluetp       & \AMetbluetpr       & \AFetblue       & \AFetbluetp       & \AFetbluetpr       \\
     & EternalSafety::EternalSynergy                 & \AMetsyn        & \AMetsyntp        & \AMetsyntpr        & \AFetsyn        & \AFetsyntp        & \AFetsyntpr        \\
     & EternalSafety::ViolationCmd                   & \AMviocmd       & \AMviocmdtp       & \AMviocmdtpr       & \AFviocmd       & \AFviocmdtp       & \AFviocmdtpr       \\
     & EternalSafety::ViolationNtRename              & \AMviontrename  & \AMviontrenametp  & \AMviontrenametpr  & \AFviontrename  & \AFviontrenametp  & \AFviontrenametpr  \\
     & EternalSafety::ViolationPidMid                & \AMviopidmid    & \AMviopidmidtp    & \AMviopidmidtpr    & \AFviopidmid    & \AFviopidmidtp    & \AFviopidmidtpr    \\
     & EternalSafety::ViolationTx2Cmd                & \AMviotxcmd     & \AMviotxcmdtp     & \AMviotxcmdtpr     & \AFviotxcmd     & \AFviotxcmdtp     & \AFviotxcmdtpr     \\
     & Signatures::Sensitive\_Signature              & \AMsensitivesig & \AMsensitivesigtp & \AMsensitivesigtpr & \AFsensitivesig & \AFsensitivesigtp & \AFsensitivesigtpr \\
    \midrule
           & Total                                            & \AMalerts       & \AMtotaltp        & \AMtotaltpr        & \AFalerts       & \AFtotaltp        & \AFtotaltpr        \\
    \bottomrule
    \end{tabular}
    \end{adjustbox}
\end{table*}

The results from this experiment indicate that our approach for APT contextualization is able to
achieve a \emph{significant reduction} in alert volume up to \IFreductionpc\% of the original alert
set. While the absolute numbers of APT scenario graphs per day, that are false positives in this
case, is not negligible with \IMscenariosperday\ and \IFscenariosperday\ it is not unreasonable to
assume that a SOC would be able to handle these numbers via analysts especially in high-security
contexts.
Additionally, it is important to note that our contextualization so far is based on network data
only. Additional IoCs,\eg{}obtained from host IDSs or sensors, can be leveraged to further filter
the set of APT scenario graphs and to offer additional context to prioritize graphs for manual
analysis.

\subsubsection*{APT campaign detection \& characterization}\label{subsubsec:results-apt}
In the second experiment, we evaluate the detection performance of our approach on the synthetic
APT campaign described above, that we called IDS2018-APT. \cref{tb:eval-overview-apt} summarizes
results of the contextualization process for both Zeek configurations. The final numbers of
\emph{APT scenario graphs} with \AMdistinctscenarios\ (MIN) and \AFdistinctscenarios\ (FULL)
respectively result in volume reductions similar to the unmodified data set.
To evaluate the functional correctness of our contextualization approach, we evaluate the
groundtruth starting from the alert level up to the generated APT scenario graphs.

\begin{table*}
    \centering
    \caption{IDS2018-APT: Groundtruth after clustering via GAC~\cite{ac-haas-2018}}\label{tb:eval-apt-meta-alerts}
    \begin{tabular}[b]{lcrrcrr}
    \toprule
                                & \multicolumn{3}{c}{IDS2018-APT-MIN}        & \multicolumn{3}{c}{IDS2018-APT-FULL} \\
                                \cmidrule(r){2-4}                            \cmidrule(r){5-7}
    Attack Step                 & Type   & Source IP(s)  & Destination IP(s) & Type   & Source IP(s)  & Destination IP(s) \\
    \midrule
    \emph{EternalRomance} RCE   & ---    & ---           & ---               & Meta   & 1.1.13.37     & \multirow{2}{*}{172.31.64.67}\\
                                &        &               &                   &        & + 7 unrelated &  \\
    2nd stage trojan download   & Single & 172.31.64.67  & 12.34.12.34       & Single & 172.31.64.67  & 12.34.12.34\\
    Cosmic Duke C\&C            & ---    & ---           & ---               & ---    & ---           & ---\\
    PS-EXEC via SMB             & Single & 172.31.64.67  & 172.31.69.20      & Single & 172.31.64.67  & 172.31.69.20\\
    Data exfiltration via HTTPS & Single & 172.31.69.20  & 1.1.15.57         & Single & 172.31.69.20  & 1.1.15.57\\
    \bottomrule
    \end{tabular}
\end{table*}

\paragraph*{Zeek alerts}
\cref{tb:eval-apt-zeek} shows all alerts generated by Zeek for the two different configurations
grouped by their \emph{alert type}. The first block contains alerts generated by scripts that were
shipped with Zeek. The second block contains all scripts written by us tailored to the organization
scenario (prefixed with \enquote{Org}). The third and fourth blocks are only included in the FULL
configuration and contain alerts produced by \emph{Mitre's BZAR} (prefixed with \enquote{ATTACK})
and \emph{EternalSafety} (prefixed with that name) respectively.
For each alert type and configuration, the table lists lists the total number of alerts observed of
that type, the total number of alerts that were produced as a result of our injected APT campaign,
and the ratio between all alerts of that type and the ones related to the campaign. This ratio is
the true positive rate (TPR) related to our APT campaign. The remaining alerts are the result of
either attacks unrelated to the APT (as present in the original data set) or false positives.
The table shows that the overall ratio of alerts related to the APT campaign is very small with
\AMtotaltprpc\% for MIN and \AFtotaltprpc\% for FULL, respectively. The type of alerts provides
some insight about which parts of our APT campaign were detected. The organization-specific script
related to exectuable downloads over a non-HTTP channel revealed the second stage trojan download
on day 1, while the SMB script detected the PS-EXEC lateral movement on day 8. The alert
\texttt{Org::Very\_Large\_Outgoing\_Tx} was generated for the exfiltration on the last day. Overall
four out of the \AMalerts\ total alerts produced by the MIN configuration relate to IDS2018-APT in
some way.
In the FULL configuration, \emph{BZAR} adds seven additional APT related alerts that are all caused
by the lateral movement. \emph{EternalSafety} produces three more APT related alerts as a result of
the initial \emph{EternalRomance} remote code execution (RCE) on day 1. While the alerts
\texttt{EternalSafety::DoublePulsar} and \texttt{EternalSafety::EternalSynergy} are all related to
our APT campaign, the other types and \texttt{EternalSafety::ViolationTx2Cmd} produce high volumes
of unrelated alerts and are responsible for a large increase of alerts from MIN to FULL.

Summing up, the MIN configuration produced Zeek alerts for the second stage trojan download, the
lateral movement via PS-EXEC, and the data exfiltration. It missed the initial RCE and the Cosmic
Duke C\&C communication, which thus cannot be contained in any further processing. The FULL
configuration improves the detection of lateral movement via additional alerts from BZAR, while
the EternalSynergy packages is able to produce a few alerts related to the initial RCE among a
large number of false positives.

\begin{figure*}
    \subfloat[IDS2018-APT-MIN\label{fig:result-synth1}]{%
        \includegraphics[width=.4\linewidth]{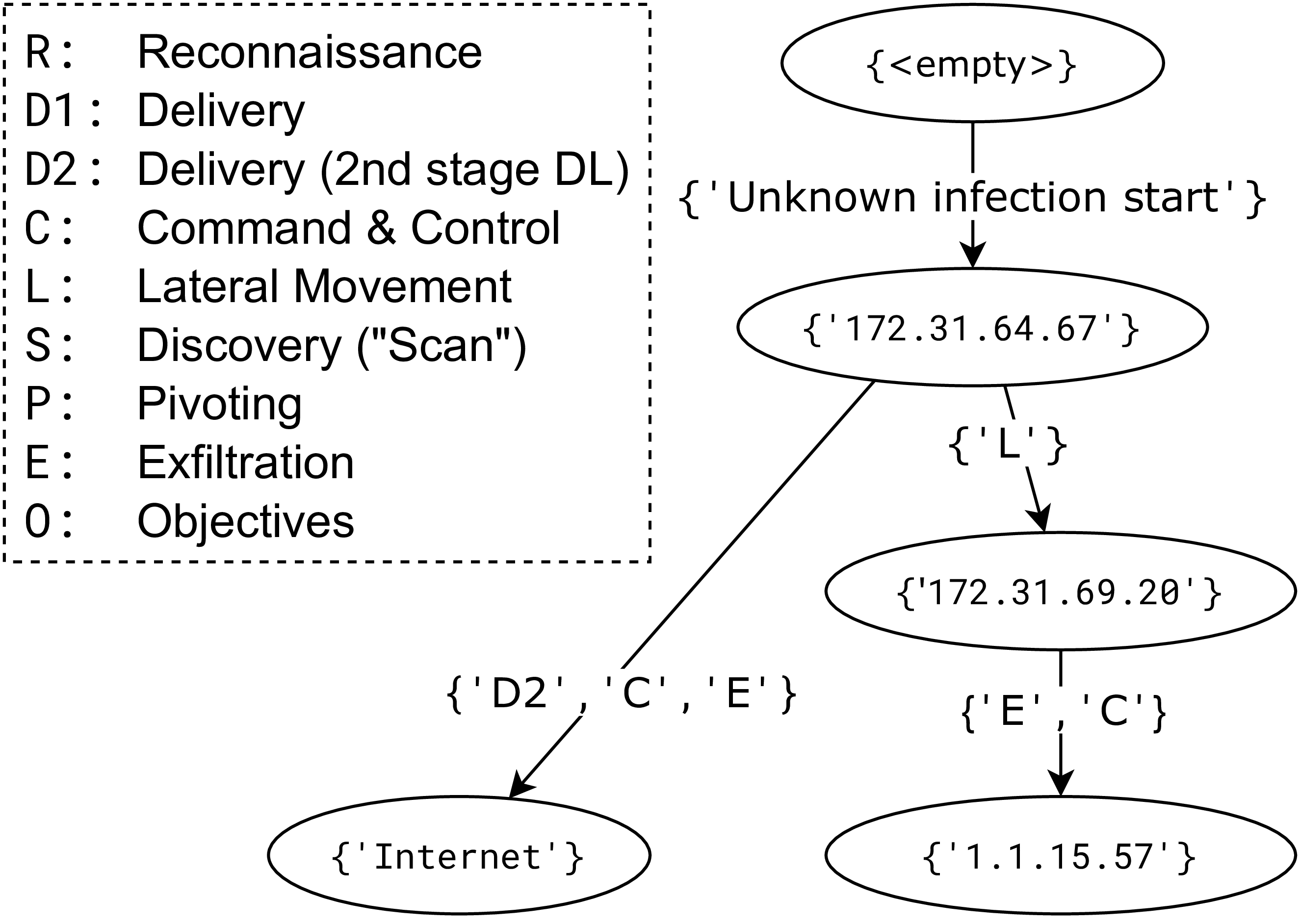}
    }
    \hspace{4em}
    \subfloat[IDS2018-APT-FULL\label{fig:result-synth1-bzar}]{%
        \includegraphics[width=.32\linewidth]{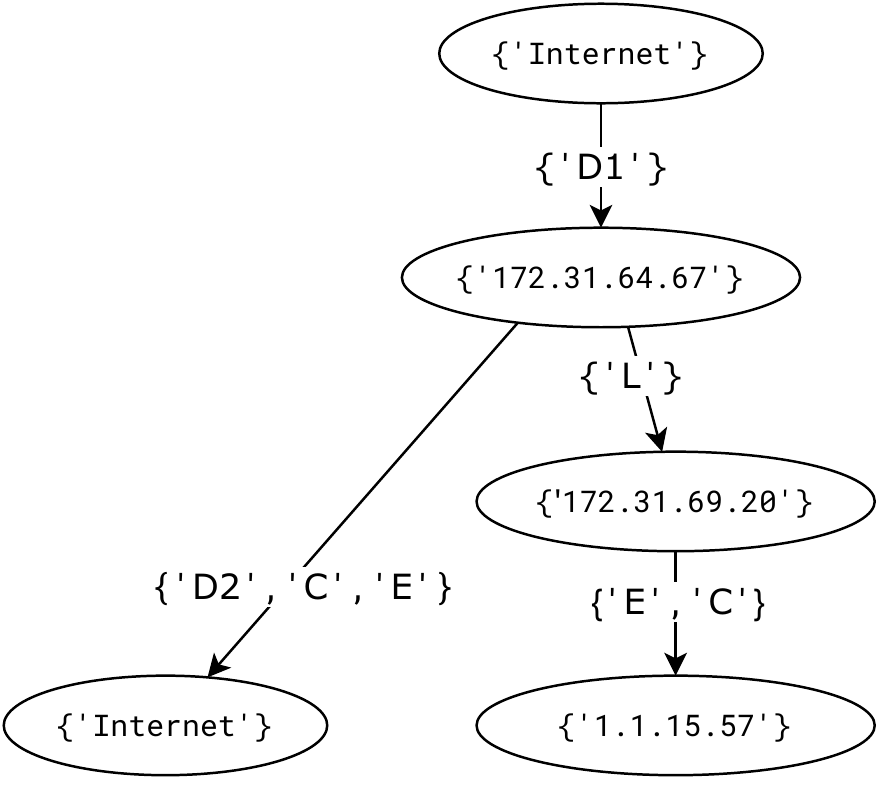}
    }
    \caption{IDS2018-APT: Resulting APT scenario graphs.}\label{fig:result-synth1-both}
\end{figure*}

\paragraph*{Meta-alerts and single alerts}
After correlation via GAC, we obtained clustered meta-alerts and unclustered singleton that both
serve as input for the APT contextualization. \cref{tb:eval-apt-meta-alerts} shows which
meta-alerts and singleton alerts are related to IDS2018-APT for both configurations.
For the MIN configuration GAC produced \AMgacalerts\ meta-alerts and \AMweakalerts\ unclustered
alerts. Out of these total \AMhighlevelalerts\ alerts, only three singleton alerts are
related to IDS2018-APT. This reinforces the expectation, that alerts caused by APT activity are not
clustered by traditional alert correlation algorithms. Given these results we expect an accurate
contextualization result as all alerts contain single IP addresses for as source and destination.
In the FULL configuration GAC generated \AFgacalerts\ meta-alerts and \AFweakalerts\ singleton
alerts. Compared to the unmodified data set these numbers are in the expected range. Out of the
four relevant alerts, we see three singleton alerts for the same attack steps as in MIN plus an
additional alert for the initial RCE as expected from the Zeek alerts show previously. However, the
alert that captured this step is a meta-alert clustered by GAC and contains multiple source IP
addresses. This has two implications: First, the resulting APT scenario graph will likely contain
all eight IP addresses as the algorithm cannot split this set without additional intelligence.
Second, this example shows that meta-alerts may also carry (partial) information related to APTs.
While the results from the MIN configuration may indicate, that it is sufficient to only
investigate unclustered singleton alerts, this does not hold in the FULL configuration. In summary,
we expect a similar contextualization result compared to MIN with more details about the initial
point of infection.

\paragraph*{APT scenario graphs}
For the MIN configuration, the APT Contextualization yielded \AMdistinctscenarios\ distinct APT
scenario graphs. From the original set of \AMalerts\ alerts, this implies a reduction to
\textbf{\AMreductionpc\%}. Among the result set is one graph that describes our APT campaign.
\cref{fig:result-synth1} shows the APT scenario graph that was generated from the alerts of
IDS2018-APT. As mentioned before the initial infection via \emph{EternalRomance} and the Cosmic
Duke C\&C did not generate an alert and are thus not included in the graph. Most stages are labeled
correctly with IP addresses and APT stages. The only imprecision relates to the second stage trojan
download, as the edge is labeled ambiguously with \texttt{[E, D2, C]} and the corresponding node
not only contains the target IP address \(12.34.12.34\) but the label `Internet'. The stage
mismatch is expected as our approach derives potential stages from network direction and can not
reduce the set of stages further without additional context for outgoing connections to the
Internet. The label is the result of imperfect meta-alert generation as the algorithm grouped
multiple Internet IP addresses and thus produced that label. However, the scenario references all
used meta-alerts and single alerts via a unique identifier.
The MAX configuration produced an alert for the initial RCE that was missed in MIN. Ideally, the
produced APT scenario graph should thus correctly identify the first node in the chain that is
labeled as with \texttt{\{<empty>\}} in \cref{fig:result-synth1}. The contextualization yielded
\AFdistinctscenarios\ distinct APT scenario graphs---a reduction to \AFreductionpc\% of the total
alert set. \cref{fig:result-synth1-bzar} shows the one that closely resembles our APT campaign.

The APT scenario graph matches our campaign with similar precision than the first. The initial RCE
is picked up and added to the scenario. However, as there were multiple alerts related to incoming
connections to \(172.31.64.67\), the node is labeled with \enquote{Internet}. The referenced alerts
indeed contain the one related to the RCE. Overall the FULL scenario matches the campaign more
closely as expected from the generated alerts. A human analyst could now use this graph to further
investigate the referenced alerts and underlying connections to reveal the APT campaign.

In summary our evaluations indicates that our approach works as intended. First, it significantly
reduces the overall set of alerts to be investigated by two to three orders of magnitude and
second, it can detect and contextualize complex attacks. Although not all nodes are labeled
perfectly, large parts of our APT campaign are visualized. While our APT campaign is obviously not
representative of all potential APT campaigns it does include zone movement that is characteristic
of typical multi-stage attacks. The movement of the attacker across hosts in different zones should
result in longer chains in our APT scenario graphs and thus reveal the campaign.

\section{Conclusion}\label{sec:conclusion}
Security operations teams must make sense of huge amounts of log and alert data to find attacks
before they become major breaches. But the high volume and false positive rate can quickly cause
alert fatigue, resulting in important attacker signals drowning in the noise.
In this paper we present a method to substantially reduce the alert volume using alert correlation
and attack contextualization. Instead of sifting through hundreds of thousands singleton alerts,
analysts can now triage incidents using multi-stage scenario graphs produced by our algorithm. We
achieve this by building a \emph{kill chain state machine~(KCSM)} that operates on clustered alert data to
identify states and transitions of multi-stage attacks. The resulting APT scenario graphs visualize
potential APT campaigns in the network and provide actionable context during investigations. Our
algorithm uses correlated meta-alerts as well as unclustered single alerts to construct \emph{APT
scenario graphs} that offer context about potential multi-stage attacks in the network to human
analysts.

We evaluated our contextualization approach on the CSE-CIC-IDS2018 data
set~\cite{ids-sharafaldin-2018} to quantify the operational overhead for real-world scenarios in
two different IDS configurations. For the minimal configuration our approach generated
\IMdistinctscenarios\ from \IMalerts\ alerts in the ten-day period, a reduction to
\textbf{\IMreductionpc\%}. In the high-security configuration, the algorithm reduced \IFalerts\
alerts to \IFdistinctscenarios, achieving a reduction to \textbf{\IFreductionpc\%}. This two to
three orders of magnitude reduction brings the data into a range that is feasible for human
analysts to process.
Furthermore, we designed a custom APT campaign on pcap level comprised of exploits that have been
actively used in complex attacks and evaluated the contextualization performance of our approach.
For both IDS configurations, our approach is able to achieve a comparable reduction as in the
unmodified data set while adding additional context about infected IP hosts and potential stages
between them. This information helps threat hunters to prioritize hosts in further investigations.
In both configurations, the algorithm produced a graph that contains all parts of our scenario that
were detectable,\ie{}an alerts was present at the lowest level.
Our approach is largely based on network direction and derives potential APT stages from the
movement between network zones. As attackers generally need to traverse multiple network zones in
any enterprise network to reach their targets, we are confident, that the algorithm is able to
track large parts of such campaigns as long as any alerts are generated. The stage deduction based
on network direction makes the approach very flexible as it does not require special information in
the underlying alerts and thus can be applied to any network based alert. Furthermore, our
algorithm can also process additional stage-specific alerts and incorporate the resulting attack
stages into the APT scenario graphs.

While our results are already quite promising, the number of potential scenarios can be further
improved. Our algorithm currently only processes network-level information. In the future, we plan
to integrate host-level as well as user identity context for richer scenario graphs and more
opportunities to eliminate false positives.

\balance%
\bibliographystyle{ACM-Reference-Format}
\bibliography{bib}


\end{document}